\documentclass[a4paper,12pt]{article}
\usepackage[a4paper,text={16.8cm,22.4cm}]{geometry}
\usepackage{amsmath,amssymb,bm,psfrag,graphicx}
\allowdisplaybreaks 
\begin{document}

\begin{titlepage}

\begin{flushright}
HD-THEP/10-13\\
MZ-TH/10-26\\
July 22, 2010
\end{flushright}

\vspace{0.2cm}
\begin{center}
\Large\bf
Drell-Yan production at small $\bm{q_T}$, transverse parton distributions and the collinear anomaly
\end{center}

\vspace{0.2cm}
\begin{center}
Thomas Becher$^a$ and Matthias Neubert$^{b,c}$\\
\vspace{0.4cm}
{\sl 
${}^a$\,Institut f\"ur Theoretische Physik, Universit\"at Bern\\
Sidlerstrasse 5, CH--3012 Bern, Switzerland\\[0.3cm]
${}^b$\,Institut f\"ur Physik (THEP), 
Johannes Gutenberg-Universit\"at\\ 
D--55099 Mainz, Germany\\[0.3cm]
${}^c$\,Institut f\"ur Theoretische Physik, Ruprecht-Karls-Universit\"at Heidelberg\\
Philosophenweg 16, D-69120 Heidelberg, Germany}
\end{center}

\vspace{0.2cm}
\begin{abstract}
\vspace{0.2cm}
\noindent 
Using methods from effective field theory, an exact all-order expression for the Drell-Yan cross section at small transverse momentum is derived directly in $q_T$ space, in which all large logarithms are resummed. The anomalous dimensions and matching coefficients necessary for resummation at NNLL order are given explicitly. The precise relation between our result and the Collins-Soper-Sterman formula is discussed, and as a by-product the previously unknown three-loop coefficient $A^{(3)}$ is obtained. The naive factorization of the cross section at small transverse momentum is broken by a collinear anomaly, which prevents a process-independent definition of $x_T$-dependent parton distribution functions. A factorization theorem is derived for the product of two such functions, in which the dependence on the hard momentum transfer is separated out. The remainder factors into a product of two functions of longitudinal momentum variables and $x_T^2$, whose renormalization-group evolution is derived and solved in closed form. The matching of these functions at small $x_T$ onto standard parton distributions is calculated at ${\cal O}(\alpha_s)$, while their anomalous dimensions are known to three loops.
\end{abstract}
\vfil

\end{titlepage}

\section{Introduction}

In collider processes with several disparate scales, fixed-order perturbative results become unreliable since higher-order corrections are enhanced by large logarithms of scale ratios. The classic example of such a multi-scale process is the production of electroweak bosons with transverse momentum $q_T$ small compared to their mass $M$. The leading logarithmically-enhanced corrections in this kinematic region were resummed in \cite{DDT,Parisi:1979se,Curci:1979bg}. An all-order formula for the cross section at small $q_T$ was obtained by Collins, Soper, and Sterman (CSS) \cite{Collins:1984kg}, and explicit results for the ingredients necessary for resummation at next-to-next-to-leading logarithmic (NNLL) order were derived in \cite{Davies:1984hs,Davies:1984sp,deFlorian:2001zd}. The region of small $q_T$ is of phenomenological importance, since it has the largest cross section and is used to extract the $W$-boson mass and width. In fact, the measurement of the charged-lepton $q_T$ spectrum now gives the most precise determination of $M_W$ \cite{:2007ypa,Aaltonen:2007ps,Abazov:2009cp}. 

While the vector-boson $q_T$ spectrum is a classical example of an observable which exhibits logarithmic enhancements, analyzing its factorization properties is nevertheless rather subtle. In Section~\ref{sec: factorization}, we study the factorization properties of the Drell-Yan cross section at low transverse momentum in Soft-Collinear Effective Theory (SCET) \cite{Bauer:2000yr,Bauer:2001yt,Beneke:2002ph}. At low $q_T$, the cross section splits into two transverse-position dependent parton distribution functions (to which we will sometimes simply refer as ``transverse PDFs'') multiplying a hard function depending only on the vector-boson mass $M$. However, it is well known that a naive definition of transverse-position dependent PDFs leads to inconsistencies due to rapidity divergences \cite{Collins:2003fm}. A proper definition requires introducing additional regulators beyond dimensional regularization. After taking the product of the two transverse PDFs these regulators can be removed, but finite terms depending on the hard momentum transfer $q^2=M^2$ of the scattering process remain. We use the term ``collinear anomaly'' for this effect, since it describes a situation where a property of the classical theory cannot be maintained after including quantum corrections. We show that the anomalous terms have a very specific structure, which allows us to exponentiate the $q^2$-dependent pieces in the product of two transverse PDFs. Once this is done factorization is restored. Specifically, we derive a factorization formula, which at fixed transverse displacement $x_T$ expresses the product of two transverse PDFs in terms of three functions capturing the dependences on the hard scale $q^2$ and the two light-cone momentum fractions $\xi_1$ and $\xi_2$. The renormalization-group (RG) equations for these functions are derived and solved in closed form. This leads us to a consistent definition of the concept of $x_T$-dependent PDFs, valid to all orders of perturbation theory. A simpler example for the occurrence of the collinear anomaly is the Sudakov form factor of a massive vector boson, which was discussed in SCET in \cite{Chiu:2007yn}. The exponentiation of the associated $\ln M^2$ dependence was demonstrated in \cite{Chiu:2007dg}. 

In the region $q_T\gg\Lambda_{\rm QCD}$, the transverse PDFs can be expanded in powers of $x_T^2$. One then recovers the standard PDFs convoluted with $x_T$-dependent kernel functions, which we compute at one-loop order in Section~\ref{sec:oneloop}. A similar situation was studied in \cite{Stewart:2009yx,Stewart:2010qs}. Instead of the transverse PDFs relevant here, these papers considered a $q_+$-dependent PDF, where $q_+$ is the small light-cone component of the partons emitted from the proton. What makes our case more complicated is the presence of the collinear anomaly, which neccessitates that both beam nucleons are considered at the same time and that an additional regulator is introduced. In our computations we use analytic regularization of the collinear propagators as in \cite{Smirnov:1993ta,Smirnov:2002pj}. We show that in $x_T$ space the $\ln q^2$ dependence induced by the collinear anomaly exponentiates to all orders in perturbation theory. In Section~\ref{sec:resum} we solve RG equations in SCET to derive, for the first time, a closed formula for the cross section in the region $\Lambda_{\rm QCD}\ll q_T\ll M$ directly in momentum ($q_T$) space, which is free of large perturbative logarithms. This formula is the analog of the classical CSS formula, which is written in impact-parameter space, and it is valid to all orders of perturbation theory. 

Besides large logarithms, the perturbative series for the transverse-momentum distribution contains terms featuring a strong factorial growth at higher orders in $\alpha_s$, which must be summed to all orders to obtain a reliable result. In Section~\ref{sec:asymp} we explain the origin of these terms and show how their summation can be implemented in closed form. In contrast to the CSS approach, our result for the resummed hard-scattering kernels is free of Landau-pole ambiguities, because we never perform scale setting inside integrals over the running coupling. Its connection to the CSS formula is discussed in Section~\ref{sec:comp}. There we also derive all ingredients necessary for NNLL resummation, in particular also the previously unknown three-loop coefficient $A^{(3)}$, and show that it is not equal to the three-loop cusp anomalous dimension, as is often assumed in the literature (see e.g.\ \cite{Bozzi:2003jy,Balazs:2007hr}). 

In previous analyses of $q_T$ resummation in SCET \cite{Gao:2005iu,Idilbi:2005er,Mantry:2009qz}, which we review in Section~\ref{sec:comp}, the collinear anomaly was regulated by keeping power-suppressed terms in the Lagrangian for the computation of the leading-power cross section. This complicates calculations, since then the various component functions no longer have homogeneous scaling in the SCET expansion parameter. More importantly, none of the previous works has addressed the resummation of the logarithms of $q_T^2/M^2$, which arise in the matching of the transverse-position dependent PDFs onto the standard PDFs. These define a class of next-to-leading logarithms, which cannot be resummed using the evolution of the hard matching coefficient. In the present work we show how these logarithms can be resummed to all orders in perturbation theory, working directly in momentum space using the formalism of \cite{Becher:2006nr}. Our work thus provides the first complete resummation of the large logarithms in the $q_T$ spectrum from effective field theory. A summary of our main results and conclusions are given in Section~\ref{sec:con}. Phenomenological applications of our approach will be presented elsewhere.

\section{Derivation of the factorization formula} 
\label{sec: factorization}

We study the Drell-Yan production of lepton pairs or electroweak gauge bosons with invariant mass $M$ and transverse momentum $q_T$ in the kinematical region where $M^2\gg q_T^2$. Here $q_\perp^\mu$ denotes the transverse-momentum vector of the boson orthogonal to the beam axis, and we denote $q_T^2\equiv-q_\perp^2\ge 0$. Our goal is to systematically resum large logarithms of the ratio $M^2/q_T^2$ to all orders in perturbation theory. Most of our analysis will assume that $q_T^2\gg\Lambda_{\rm QCD}^2$ is in the perturbative domain, but we will also discuss the case where $q_T^2\sim\Lambda_{\rm QCD}^2$ in some detail. 

\subsection{Kinematical considerations}

For concreteness we consider the production of a lepton pair via a virtual photon with total momentum $q^\mu$. The cross sections for Drell-Yan production of $W$ and $Z$ bosons can be obtained from the results presented here by means of simple substitutions summarized in Appendix~\ref{app:b}. We begin with the standard relation 
\begin{equation}\label{dsig}
   d\sigma = \frac{4\pi\alpha^2}{3 q^2 s}\,\frac{d^4q}{(2\pi)^4} 
    \int\!d^4x\,e^{-iq\cdot x}\,(-g_{\mu\nu})\,
    \langle N_1(p)\,N_2(\bar p)|\,J^{\mu\dagger}(x)\,J^\nu(0)\,
    |N_1(p)\,N_2(\bar p)\rangle \,,
\end{equation}
where 
\begin{equation}\label{Jmu}
   J^\mu = \sum_q \left( g_L^q\,\bar q\gamma^\mu\,\frac{1-\gamma_5}{2}\,q
    + g_R^q\,\bar q\gamma^\mu\,\frac{1+\gamma_5}{2}\,q \right) ,
    \qquad
   g_L^q = g_R^q = e_q
\end{equation}
is the electromagnetic current. Keeping the left- and right-handed couplings separate has the advantage that our analysis can be carried over straightforwardly to the case of weak gauge-boson production.

To analyze this process in SCET, we introduce two light-like reference vectors $n$ and $\bar n$ satisfying $n\cdot\bar n=2$, which are parallel to the directions of the colliding hadrons $N_1$ and $N_2$ with momenta $p$ and $\bar p$. Any four momentum can then be split into light-cone and perpendicular components according to
\begin{equation}
   k^\mu = n\cdot k\,\frac{\bar n^\mu}{2} + \bar n\cdot k\,\frac{n^\mu}{2} 
    + k_\perp^\mu 
   \equiv k_+^\mu + k_-^\mu + k_\perp^\mu \,.
\end{equation}
In the effective theory one defines a small expansion parameter $\lambda=q_T/M$ and distinguishes fields whose momentum components $(n\cdot k,\bar n\cdot k, k_\perp)$ scale differently with $\lambda$. In the present case, the production of a lepton pair with transverse momentum $q_T$ requires one or more parton emissions into the final state, which balance that momentum. As a result, the substructure of the colliding hadrons is probed at distance scales of order $1/q_T$. The partons in the colliding beam jets thus generically have so-called hard-collinear ($hc$) or anti-hard-collinear ($\overline{hc}$) momenta scaling as
\begin{equation}
   p_{hc}\sim M \,(\lambda^2,1,\lambda) \,, \qquad
   p_{\overline{hc}}\sim M\,(1,\lambda^2,\lambda) \,.
\end{equation}
In SCET, one introduces different sets of hard-collinear and anti-hard-collinear quark and gluon fields describing the interactions of these partons \cite{Bauer:2000yr,Bauer:2001yt,Beneke:2002ph}. While these fields have QCD-like interactions among themselves, two fields belonging to different sectors can interact only via the exchange of soft partons (only soft gluon interactions contribute at leading power in $\lambda$), whose momenta scale like
\begin{equation}
   p_s\sim M \,(\lambda^2,\lambda^2,\lambda^2) \,.
\end{equation}
Adding a soft momentum to a (anti-)hard-collinear momentum does not change its scaling properties. Since in our case the total transverse momentum of the hadronic final state must balance the transverse momentum $q_T=M\lambda$ of the lepton pair, it follows that this final state must contain at least one (anti-)hard-collinear parton, and that soft fields give a power-suppressed contribution to its transverse momentum. As a result, we will see that the total contribution of an arbitrary number of soft emissions cancels in the final factorization formula for the Drell-Yan cross section at fixed $q_T$. This is in contrast with the factorization formula for Drell-Yan production near threshold, in which soft modes give an important contribution \cite{Sterman:1986aj,Catani:1989ne} (see \cite{Becher:2007ty} for a derivation using SCET). The absence of a soft contribution is an important fact. If there were a sensitivity to the soft scale $\mu_s\sim q_T^2/M$, the Drell-Yan $q_T$ distribution would be non-perturbative even at high transverse momenta $q_T\lesssim\sqrt{M \Lambda_{\rm QCD}}$, which would be a disaster for phenomenology.

One might wonder whether also partons with momenta scaling as $M(\lambda,\lambda,\lambda)$ should be considered, which could contribute to the spectrum at leading power. On the one hand, the appearance of such a semi-hard mode would be quite unexpected from the point of view of SCET, since it is not allowed to interact with the (anti-)hard-collinear partons. Indeed, a semi-hard mode does not contribute to any other collider-physics process analyzed with SCET so far. One the other hand, because the transverse momentum of the final-state partons is restricted to scale like $M\lambda$, it is not obvious that the emission of semi-hard partons is really irrelevant. In Section~\ref{sec:oneloop}, we will show that the semi-hard region does not contribute if the collinear anomaly is regularized analytically. In this respect our analysis differs from a recent study in \cite{Mantry:2009qz}. We will comment on this paper in more detail in Section~\ref{sec:comp}.

In SCET, the current (\ref{Jmu}) is matched onto (we adopt a regularization scheme with anti-commuting $\gamma_5$) \cite{Manohar:2003vb,Becher:2006mr}
\begin{equation}\label{eq:current}
   J^\mu \to C_V(-q^2-i\varepsilon,\mu)\,\sum_q
    \left( g_L^q\,\bar\chi_{\overline{hc}}\,S^\dagger_{\bar n}\,\gamma^\mu\,
    \frac{1-\gamma_5}{2}\,S_n\,\chi_{hc}
    + g_R^q\,\bar\chi_{\overline{hc}}\,S^\dagger_{\bar n}\,\gamma^\mu\,
    \frac{1+\gamma_5}{2}\,S_n\,\chi_{hc} \right) ,
\end{equation}
where the matching coefficient $C_V$ depends on the hard momentum transfer $Q^2=-q^2=-M^2$. For time-like processes this coefficient is complex. From now on we will suppress the $-i\varepsilon$ prescription that defines the sign of the imaginary part of $C_V$. The effective fields $\chi_{hc}=W_{hc}^\dagger\,\xi_{hc}$ and $\chi_{\overline{hc}}=W_{\overline{hc}}^\dagger\,\xi_{\overline{hc}}$ are the usual gauge-invariant combinations of effective (anti-)hard-collinear quark fields and Wilson lines.\footnote{As usual in SCET we restrict ourselves to gauge transformations that vanish at infinity.} 
They satisfy $\rlap/n\,\chi_{hc}=0$ and $\rlap/\bar n\,\chi_{\overline{hc}}=0$. These fields are obtained after the SCET decoupling transformation has been applied, which removes the interactions between soft and (anti-)hard-collinear fields in the leading-order SCET Lagrangian \cite{Bauer:2001yt}. This introduces the soft Wilson lines $S_n$ and $S_{\bar n}$ into the effective current shown above. In \cite{Becher:2007ty}, we have derived a factorization theorem for the Drell-Yan process in the threshold region. Many more details about the effective-theory formalism can be found in this reference. 

Using a Fierz transformation along with some elementary Dirac algebra, and averaging over nucleon spins, the hadronic matrix element in (\ref{dsig}) can be rewritten as
\begin{equation}\label{g2formula}
\begin{aligned}
   & (-g_{\mu\nu})\,\langle N_1(p)\,N_2(\bar p)|\,J^{\mu\dagger}(x)\,J^\nu(0)\,
    |N_1(p)\,N_2(\bar p)\rangle 
    \to |C_V(-q^2,\mu)|^2 \sum_q\,\frac{|g_L^q|^2+|g_R^q|^2}{2N_c} \\[-2mm]
   &\hspace{7mm}\times \hat W_{\rm DY}(x)\,
    \langle N_1(p)|\,\bar\chi_{hc}(x)\,\frac{\rlap{/}{\bar n}}{2}\,\chi_{hc}(0)\,
    |N_1(p)\rangle\,
    \langle N_2(\bar p)|\,\bar\chi_{\overline{hc}}(0)\,\frac{\rlap{/}{n}}{2}\,
    \chi_{\overline{hc}}(x)\,|N_2(\bar p)\rangle \,,
\end{aligned}
\end{equation}
where
\begin{equation}\label{Wsoft}
   \hat W_{\rm DY}(x) = \frac{1}{N_c}\,\langle 0|\,\mbox{Tr}
   \big[ \overline{\bf T} \big( S_n^\dagger(x)\,S_{\bar n}(x) \big)\,
    {\bf T} \big( S_{\bar n}^\dagger(0)\,S_n(0) \big) \big] |0\rangle 
\end{equation}
is a soft Wilson-line correlator. Up to this point our discussion is completely analogous to that in \cite{Becher:2007ty}. We now perform the multipole expansion appropriate for the kinematical situation considered here, i.e., we expand the fields in derivatives corresponding to suppressed momentum components \cite{Beneke:2002ph,Beneke:2002ni}. Here we encounter a difference with our previous analysis, which is due to the different kinematical requirements on the transverse momentum of the photon. In the present case the separation of the fields scales as $x\sim M^{-1}(1,1,\lambda^{-1})$, which is conjugate to the hard photon momentum $q\sim M(1,1,\lambda)$. Since derivatives on the soft fields scale as $\lambda^2$, the leading term in the expansion is obtained by evaluating the soft Wilson lines at $x=0$. In the (anti-)hard-collinear fields, the dependence on $x_\perp$ and on the light-cone components conjugate to the large momentum components is unsuppressed and must be kept. The expanded result then takes the form
\begin{equation}\label{step1}
\begin{aligned}
   d\sigma &= \frac{4\pi\alpha^2}{3 N_c q^2 s}\,\frac{d^4q}{(2\pi)^4} 
    \int\!d^4x\,e^{-iq\cdot x}\,|C_V(-q^2,\mu)|^2\,\sum_q\,
    \frac{|g_L^q|^2+|g_R^q|^2}{2}\,\hat W_{\rm DY}(0) \\[-2mm]
   &\quad\times \langle N_1(p)|\,\bar\chi_{hc}(x_+ + x_\perp)\,\frac{\rlap{/}{\bar n}}{2}\,
    \chi_{hc}(0)\,|N_1(p)\rangle\,
    \langle N_2(\bar p)|\,\bar\chi_{\overline{hc}}(0)\,\frac{\rlap{/}{n}}{2}\,
    \chi_{\overline{hc}}(x_- + x_\perp)\,|N_2(\bar p)\rangle \,.
\end{aligned}
\end{equation}
The definition (\ref{Wsoft}) implies that $\hat W_{\rm DY}(0)=1$, and hence the soft contribution cancels out in the result for the cross section. Physically, the reason is simply that soft gluons carry transverse momenta of order $q_T^2/M$, which for $M\gg q_T$ implies that an arbitrary number of soft gluons can be emitted into the final state without changing the total transverse momentum $q_T$. Hence, real and virtual soft-gluon effects cancel by the KLN theorem.

We recall at this point the definitions of the parton distribution functions (PDFs) \cite{Efremov:1978xm,Collins:1981uk,Collins:1981uw}. In terms of SCET operators, they read  
\begin{equation}\label{phidef}
\begin{aligned}
   \phi_{q/N}(z,\mu) 
   &= \frac{1}{2\pi} \int dt\,e^{-izt\bar n\cdot p}\,
    \langle N(p)|\,\bar\chi(t\bar n)\,\frac{\rlap/\bar n}{2}\,\chi(0)\,|N(p)\rangle \,, \\
   \phi_{g/N}(z,\mu) 
   &= \frac{z\,\bar n\cdot p}{2\pi} \int dt\,e^{-izt\bar n\cdot p}\,
    \langle N(p)|-\!{\cal A}_{\perp\mu}(t\bar n)\,{\cal A}_\perp^\mu(0)\,|N(p)\rangle \,,    
\end{aligned}
\end{equation}
where $0\le z\le 1$, and for simplicity we suppress the $hc$ labels on the fields. The nucleon $N$ carries momentum $p$ along the $n$ direction. The anti-quark PDF $\phi_{\bar q/N}(z,\mu)$ is given by the same matrix element as in the first line, but with the sign in the exponent reversed. Because of the appearance of the transverse displacement vector $x_\perp$, the hadronic matrix elements in (\ref{step1}) are not of the form of the matrix elements of light-ray operators defining the PDFs. Instead, one defines the generalized, $x_T$-dependent PDFs (with $x_T^2\equiv -x_\perp^2>0$) \cite{Collins:1981uk,Collins:1981uw}
\begin{equation}\label{Bdef}
   {\cal B}_{q/N}(z,x_T^2,\mu) 
   = \frac{1}{2\pi} \int dt\,e^{-izt\bar n\cdot p}\,\langle N(p)|\,
    \bar\chi(t\bar n+x_\perp)\,\frac{\rlap/\bar n}{2}\,\chi(0)\,|N(p)\rangle \,,     
\end{equation}
and similarly for the gluon and anti-quark cases. Their Fourier transforms with respect to $x_T$ are referred to as transverse-momentum dependent PDFs. Naively, then, the differential cross section (\ref{step1}) can be expressed in terms of a convolution of the hard matching coefficient $|C_V(-M^2,\mu)|^2$ 
 with $x_T$-dependent PDFs,
\begin{equation}\label{Bfactnaive}
\begin{aligned}
   \frac{d^3\sigma}{dM^2\,dq_T^2\,dy} 
   &= \frac{4\pi\alpha^2}{3N_c M^2 s} \left| C_V(-M^2,\mu)\right|^2 
    \frac{1}{4\pi} \int\!d^2x_\perp\,e^{-iq_\perp\cdot x_\perp} \\
   &\quad\times \sum_q\,e_q^2\,\bigg[ {\cal B}_{q/N_1}(\xi_1,x_T^2,\mu)\,
    {\cal B}_{\bar q/N_2}(\xi_2,x_T^2,\mu) + (q\leftrightarrow\bar q) \bigg] 
    + {\cal O}\bigg( \frac{q_T^2}{M^2} \bigg) \,,
\end{aligned}
\end{equation}
where we have used that $g_L^q=g_R^q=e_q$, and have defined
\begin{equation}\label{taudef}
   \xi_1 = \sqrt{\tau}\,e^y \,, \qquad
   \xi_2 = \sqrt{\tau}\,e^{-y} \,, \qquad
   \mbox{with} \quad
   \tau = \frac{m_\perp^2}{s} = \frac{M^2+q_T^2}{s} \,.
\end{equation}
We have also employed that $d^4q\,\theta(q^0)\,\delta(q^2-M^2)=\frac12\,d^2q_\perp\,dy=\frac{\pi}{2}\,dq_T^2\,dy$, where the last identity holds after integration over the azimuthal angle. The above formula appears to achieve the desired factorization of the hard and hard-collinear scales, $M^2$ and $q_T^2\sim x_T^{-2}$.

\subsection{Collinear anomaly and refactorization}
\label{sec:collinearanomaly}

The formal derivation of factorization just presented is spoiled by quantum effects. This can be seen from the fact that the RG equation for the hard matching coefficient contains a term proportional to a ``cusp logarithm'' of the hard scale $q^2=M^2$ \cite{Becher:2006nr},
\begin{equation}\label{CV2evol}
   \frac{d}{d\ln\mu}\,C_V(-q^2,\mu) 
   = \left[ \Gamma_{\rm cusp}^F(\alpha_s)\,\ln\frac{-q^2}{\mu^2} + 2\gamma^q(\alpha_s)
    \right] C_V(-q^2,\mu) \,.
\end{equation}
Its coefficient $\Gamma_{\rm cusp}^F(\alpha_s)$ is the cusp anomalous dimension in the fundamental representation. The quantity $\gamma^q$ (equal to $\gamma^V/2$ in \cite{Becher:2006nr}) refers to the quark anomalous dimension as defined in \cite{Becher:2009cu,Becher:2009qa}. Here and below, the coupling $\alpha_s$ without an explicit scale argument always refers to $\alpha_s(\mu)$. RG invariance of the physical cross section (\ref{step1}) requires that the evolution equation for the {\em product\/} of the two transverse PDFs must contain the same $\ln(q^2/\mu^2)$ term as in (\ref{CV2evol}), but with the opposite sign. This fact is incompatible with the concept of universal (i.e., process-independent) transverse-position dependent PDFs. As we will see in Section~\ref{sec:1loop}, the $x_T$-dependent PDFs as given in (\ref{Bdef}), in which gauge invariance is ensured by means of light-like Wilson lines, are not well-defined in dimensional regularization and require an additional regularization of light-cone singularities. The anomalous $q^2$ dependence arises because the formal factorization of hard-collinear and anti-hard-collinear fields in the SCET Lagrangian in the absence of soft interactions is invalidated as soon as one introduces a regulator to give meaning to the loop integrals appearing in explicit evaluations of SCET diagrams. We will refer to this effect as the ``collinear anomaly''. Only the product of two transverse PDFs referring to hadrons moving in different directions is regularization independent. A similar phenomenon has been encountered in the context of SCET analyses of the $B\to\pi$ form factor \cite{Becher:2003qh,Beneke:2003pa,Lange:2003pk}, where however the factorization breakdown happens at non-perturbative scales.\footnote{After submission of this paper, we learned that in this context the term ``factorization anomaly'' was used in lectures delivered by M.~Beneke (http://theor.jinr.ru/~hq2005/Lectures/Beneke/Beneke-Dubna-05.pdf).}

The collinear anomaly is, of course, not a new quantum anomaly of QCD. Rather, it is a feature affecting certain matrix elements in SCET -- the {\em effective theory\/} of QCD relevant to the derivation of QCD factorization theorems. In SCET this effect is an anomaly in the usual sense that quantum corrections destroy a symmetry of the classical theory. At the classical level, the hard-collinear sector in the effective Lagrangian does not know about the anti-hard-collinear momentum $\bar p$, and it is thus invariant under the rescaling transformation $\bar p\to\bar\lambda\bar p$. Likewise, the anti-hard-collinear sector in the effective Lagrangian is invariant under the rescaling transformation $p\to\lambda p$. We will show in Section~\ref{sec:1loop} that in the presence of the regulators required to give meaning to SCET loops integrals only the subgroup of transformations with $\lambda\bar\lambda=1$ is left unbroken. While $q^2=2p\cdot\bar p$ is not invariant under independent rescalings of $p$ and $\bar p$, it {\em is\/} invariant under the subgroup $\lambda\bar\lambda=1$. This explains why an anomalous dependence on $q^2$ can arise in the regularized theory. Fortunately, the anomalous terms have a very specific structure, and this will allow us to exponentiate the $q^2$ dependence of the product of two transverse PDFs, thereby restoring factorization.

Because of the dependence of the transverse PDFs on the hard scale of the underlying process, relation (\ref{Bfactnaive}) does not accomplish a complete separation of the hard and hard-collinear scales $q^2=-M^2$ and $q_T^2$. It is thus not yet a useful factorization formula. In order to complete the factorization and resum all associated large logarithms, it is necessary to control the $q^2$ dependence of the product ${\cal B}_{q/N_1}\,{\cal B}_{\bar q/N_2}$ of transverse PDFs to all orders in perturbation theory. We will show in Section~\ref{sec:refact} that in $x_T$ space this product can be refactorized in the form
\begin{equation}\label{refact}
   \left[ {\cal B}_{q/N_1}(z_1,x_T^2,\mu)\,{\cal B}_{\bar q/N_2}(z_2,x_T^2,\mu)
    \right]_{q^2}
   = \left( \frac{x_T^2 q^2}{4 e^{-2\gamma_E}} \right)^{-F_{q\bar q}(x_T^2,\mu)}
    \!B_{q/N_1}(z_1,x_T^2,\mu)\,B_{\bar q/N_2}(z_2,x_T^2,\mu) \,,
\end{equation}
where the exponent $F_{q\bar q}$ depends only on the transverse coordinate and the renormalization scale. The bracket on the left-hand side of (\ref{refact}) indicates the hidden $q^2$ dependence induced by the collinear anomaly. The functions $B_{i/N}$ on the right-hand side are independent of the hard momentum transfer. While we do not have an operator definition of these functions, they are uniquely defined by relation (\ref{refact}). All $q^2$ dependence is now explicit and controlled by the function $F_{q\bar q}$. Note that for $\mu\sim x_T^{-1}$ the $q^2$-dependent prefactor resums all large logarithms of the hard scale, while $F_{q\bar q}(x_T^2,\mu)$ has a perturbative expansion in $\alpha_s(\mu)$ with ${\cal O}(1)$ coefficients. 

RG invariance of the cross section (\ref{step1}) implies the evolution equations 
\begin{equation}\label{Bevol}
\begin{aligned}
   \frac{dF_{q\bar q}(x_T^2,\mu)}{d\ln\mu} 
   &= 2\Gamma_{\rm cusp}^F(\alpha_s) \,, \\
   \frac{d}{d\ln\mu}\,B_{q/N}(z,x_T^2,\mu)
   &= \left[ \Gamma_{\rm cusp}^F(\alpha_s)\,\ln\frac{x_T^2\mu^2}{4 e^{-2\gamma_E}}
    - 2\gamma^q(\alpha_s) \right] B_{q/N}(z,x_T^2,\mu) \,.
\end{aligned}
\end{equation}
The first relation completely determines the scale dependence of $F_{q\bar q}$. Note that the structure of the evolution equation for $B_{i/N}$ is completely analogous to that for the hard matching coefficient in (\ref{CV2evol}). As a side remark, let us add that for gluon-initiated processes such as Higgs-boson production analogous evolution equations hold for the quantities $F_{gg}$ and $B_{g/N}$, in which $\Gamma_{\rm cusp}^F$ is replaced with the cusp anomalous dimension $\Gamma_{\rm cusp}^A$ in the adjoint representation, and $\gamma^q$ is replaced with the gluon anomalous dimension $\gamma^g$ as defined in \cite{Becher:2009cu,Becher:2009qa}. Formal all-order solutions to the evolution equations (\ref{CV2evol}) and (\ref{Bevol}) will be presented in Section~\ref{sec:resum} and Appendix~\ref{app:a}. Moreover, we will conjecture in Section~\ref{sec:refact} that the exponent $F_{q\bar q}$ in (\ref{refact}) is constrained by the non-abelian exponentiation theorem \cite{Gatheral:1983cz,Frenkel:1984pz}, which implies the Casimir-scaling relation
\begin{equation}\label{Casimir}
   \frac{F_{q\bar q}(x_T^2,\mu)}{C_F} = \frac{F_{gg}(x_T^2,\mu)}{C_A}
\end{equation}
at least to three-loop order.

From the point of view of our SCET analysis, equations~(\ref{refact})--(\ref{Casimir}) contain all there is to say about the concept of $x_T$-dependent PDFs. The factorization theorem (\ref{refact}) specifies in a precise, all-order way that the dependences on the variables $q^2$, $z_1$, and $z_2$ can be factorized {\em at fixed $x_T^2$} into three functions. The exact evolution equations for these functions are given in (\ref{Bevol}) and can be solved in closed form. Finally, relation (\ref{Casimir}) relates the exponents of the $q^2$-dependent terms in the quark-antiquark and gluon-gluon channels. We will show in the next section that for small transverse separation, $x_T^2\ll\Lambda_{\rm QCD}^{-2}$, the exponent $F_{q\bar q}$ and the functions $B_{i/N}$ can be evaluated in perturbation theory, the latter ones being related to standard PDFs in a calculable way. For larger values $x_T^2\sim\Lambda_{\rm QCD}^{-2}$, however, the functions $F_{q\bar q}$ and $B_{i/N}$ are genuine, non-perturbative objects, which must be modeled or extracted from fits to data. Relation (\ref{refact}) provides a rigorous basis for such modeling. We will discuss below how an experimental determination of the non-perturbative functions could be performed in practice.

A review of the conventional view on transverse-position dependent PDFs can be found in \cite{Collins:2003fm} (see also \cite{Cherednikov:2009aq}). There it is emphasized that these objects are indeed problematic, because they are subject to light-cone singularities not accounted for by standard renormalization methods. These singularities originate from gluons with very large rapidity relative to the colliding hadrons, and regularizing them requires some sort of a rapidity cut-off. While in the present work we use analytic regularization for this purpose, two different proposals were made in \cite{Collins:2003fm}: the introduction of Wilson lines off the light cone (and subject to certain constraints) \cite{Collins:1989bt}, and a ``generalized renormalization prescription'' consisting of multiplying the original $x_T$-dependent PDFs (\ref{Bdef}) with a gauge-invariant factor that cancels the extra divergences \cite{Collins:1999dz,Collins:2000gd}. These constructions have been explored at one-loop order only. Most practical applications of $x_T$-dependent PDFs employ a definition based on an axial gauge $v\cdot A=0$, as originally introduced by Collins and Soper in \cite{Collins:1981uk,Collins:1981uw} and denoted by $\tilde{\cal P}_{i/N}(z,x_T,\mu;\zeta)$. This introduces a dependence on a ``gauge parameter'' $\zeta=(2p\cdot v)^2/v^2$ and thereby a sensitivity to the energy $p^0$ of the external nucleon in the reference frame defined by $v^0=1$.\footnote{In the equivalent description using Wilson lines pointing in a time-like direction $v$, the gauge dependence is replaced by a dependence on the auxiliary vector $v$ (see e.g.\ \cite{Ji:2004wu}).}  
The product of the  gauge parameters associated with the two nuclei obeys $\sqrt{\zeta_1\zeta_2}=2p\cdot\bar p=q^2/(z_1 z_2)$, which is a gauge-independent result \cite{Collins:2000gd}. In this way the hard momentum transfer $q^2$ enters in the Collins-Soper approach, even though the precise way in which the $\zeta$ dependence cancels between the $x_T$-dependent PDFs and their hard function is not very transparent. The reason is that the $\zeta$ dependence of $\tilde{\cal P}_{i/N}(z,x_T,\mu;\zeta)$, as given by relation~(6.23) in \cite{Collins:1981uk}, is much more complicated that the $q^2$ dependence in our result (\ref{refact}).

Despite of these efforts, the precise concept of transverse-position dependent PDFs has remained  elusive \cite{Collins:2008ht}. The insight that only the {\em product\/} of transverse PDFs relevant in a given process is physical and can be defined in a regularization-independent way is a crucial new ingredient of our analysis. Moreover, our results (\ref{refact}) and (\ref{Bevol}) define the functions $B_{i/N}$ and $F_{q\bar q}$ and their evolution equations even at a non-perturbative level. On the other hand, it is an interesting open question whether the functions $B_{i/N}$ are universal and appear also in other processes (see \cite{Collins:2004nx} for a related discussion). It seems conceivable to us that in more complicated processes involving more than two directions of large energy flow in the hard-scattering event, relation (\ref{refact}) could take a more complicated form.

In terms of the objects defined in (\ref{refact}), the differential Drell-Yan cross section at fixed invariant mass $M$, transverse momentum $q_T$, and rapidity $y$ of the Drell-Yan pair can be expressed as
\begin{equation}\label{Bfact}
\begin{aligned}
   \frac{d^3\sigma}{dM^2\,dq_T^2\,dy} 
   &= \frac{4\pi\alpha^2}{3N_c M^2 s} \left| C_V(-M^2,\mu)\right|^2 
    \frac{1}{4\pi} \int\!d^2x_\perp\,e^{-iq_\perp\cdot x_\perp} 
    \left( \frac{x_T^2 M^2}{4e^{-2\gamma_E}} \right)^{-F_{q\bar q}(x_T^2,\mu)} \\
   &\quad\times \sum_q\,e_q^2\,\bigg[ B_{q/N_1}(\xi_1,x_T^2,\mu)\,
    B_{\bar q/N_2}(\xi_2,x_T^2,\mu) + (q\leftrightarrow\bar q) \bigg] 
    + {\cal O}\bigg( \frac{q_T^2}{M^2} \bigg) \,.
\end{aligned}
\end{equation}
The disparate scales $M^2$ and $q_T^2\sim x_T^{-2}$ are now completely separated, and all large logarithms can be resummed by choosing $\mu\sim q_T$ (or $\mu$ equal to a few GeV in the case where $q_T\sim\Lambda_{\rm QCD}$) and employing the RG-improved expression for $C_V(-M^2,\mu)$ given in relation (\ref{CVsol}) below. Corrections to the leading term in the factorization formula are suppressed by powers of the ratio $q_T^2/M^2\ll 1$. Also, as written above, the formula holds irrespective of whether or not the transverse momentum is a perturbative scale. Taking a Fourier transform of the cross section, it is possible to get direct access to the $x_T$-dependent PDFs as given in the factorization theorem (\ref{refact}). We find
\begin{equation}\label{invert}
\begin{aligned}
   \frac{9M^2 s}{4\pi\alpha^2} 
   &\int_0^\infty\!dq_T^2\,J_0(q_T x_T)\,\frac{d^3\sigma}{dM^2\,dq_T^2\,dy} 
   = \left| C_V(-M^2,\mu)\right|^2
    \left( \frac{x_T^2 M^2}{4e^{-2\gamma_E}} \right)^{-F_{q\bar q}(x_T^2,\mu)} \\
   &\times \sum_q\,e_q^2\,\bigg[ B_{q/N_1}(\xi_1,x_T^2,\mu)\,
    B_{\bar q/N_2}(\xi_2,x_T^2,\mu) + (q\leftrightarrow\bar q) \bigg] 
    + {\cal O}\bigg( \frac{1}{x_T^2 M^2} \bigg) \,.
\end{aligned}
\end{equation}
By varying $x_T$, $M^2$, $s$, $y$, and the beam nuclei $N_1$, $N_2$, one can (at least in principle) map out the functional dependences of $F_{q\bar q}$ and certain combinations of transverse PDFs on $x_T^2$ and $\xi_i$, much in the same way as the standard PDFs are constrained from fits to Drell-Yan cross sections. While for $x_T\ll\Lambda_{\rm QCD}^{-1}$ the right-hand side of (\ref{invert}) can be calculated in terms of known PDFs (see below), for $x_T\sim\Lambda_{\rm QCD}^{-1}$ the above relation provides access to the non-perturbative behavior of $F_{q\bar q}$ and of the transverse PDFs. This can help to constrain phenomenological models of these functions, which are needed e.g.\ for a precision determination of the mass of the $W$ boson. We emphasize that the above relation only holds for $x_T^2\gg 1/M^2$, because otherwise the power corrections to our factorization formula become large. It can therefore not be used to study the $x_T\to 0$ limit of the functions $F_{q\bar q}$ or $B_{i/N}$.

\subsection{Simplifications at large $\bm{q_T^2}$}

For given transverse momentum $q_T$, the Fourier integral in (\ref{Bfact}) receives important contributions from transverse separations $x_T\lesssim q_T^{-1}$ only. For large transverse momenta in the perturbative domain, $q_T^2\gg\Lambda_{\rm QCD}^2$, we therefore need the $x_T$-dependent PDFs at transverse separation $x_T\ll\Lambda_{\rm QCD}^{-1}$. In this case these functions obey an operator-product expansion of the form \cite{Collins:1984kg,Collins:1981uk,Collins:1981uw}
\begin{equation}\label{OPE}
   B_{i/N}(\xi,x_T^2,\mu) 
   = \sum_j \int_\xi^1\!\frac{dz}{z}\,I_{i\leftarrow j}(z,x_T^2,\mu)\,
    \phi_{j/N}(\xi/z,\mu) + {\cal O}(\Lambda_{\rm QCD}^2\,x_T^2) \,.     
\end{equation}
In the context of SCET, generalized PDFs defined in terms of hadron matrix elements in which collinear fields are separated by distances that are not light-like are referred to as beam functions. For such functions an analogous expansion was considered in \cite{Stewart:2009yx}, and an expression for the one-loop kernel of the quark beam function was derived in \cite{Stewart:2010qs}. The evolution equations for the new kernels $I_{i\leftarrow j}$ follow when we combine (\ref{Bevol}) with the standard DGLAP equations
\begin{equation}
   \frac{d}{d\ln\mu}\,\phi_{i/N}(z,\mu)
   = \sum_j \int_z^1\!\frac{du}{u}\,{\cal P}_{i\leftarrow j}(z/u,\mu)\,\phi_{j/N}(u,\mu) \,.
\end{equation}
We obtain
\begin{equation}\label{Ievol}
\begin{aligned}
   \frac{d}{d\ln\mu}\,I_{q\leftarrow i}(z,x_T^2,\mu)
   &= \left[ \Gamma_{\rm cusp}^F(\alpha_s)\,\ln\frac{x_T^2\mu^2}{4 e^{-2\gamma_E}}
    - 2\gamma^q(\alpha_s) \right] I_{q\leftarrow i}(z,x_T^2,\mu) \\
   &\quad\mbox{}- \sum_j \int_z^1\!\frac{du}{u}\,I_{q\leftarrow j}(u,x_T^2,\mu)\, 
    {\cal P}_{j\leftarrow i}(z/u,\mu) \,.
\end{aligned}
\end{equation}
Because of the complicated form of the DGLAP equations, no closed solution can be derived.

Neglecting power corrections of order $\Lambda_{\rm QCD}^2/q_T^2$, we can use relation (\ref{OPE}) to express the differential cross section (\ref{Bfact}) as a convolution of perturbative, factorized hard-scattering kernels 
\begin{equation}\label{Cdef}
\begin{aligned}
   C_{q\bar q\to ij}(z_1,z_2,q_T^2,M^2,\mu)
   &= \left| C_V(-M^2,\mu) \right|^2
    \frac{1}{4\pi} \int\!d^2x_\perp\,e^{-iq_\perp\cdot x_\perp}
    \left( \frac{x_T^2 M^2}{4e^{-2\gamma_E}} \right)^{-F_{q\bar q}(x_T^2,\mu)} \\
   &\quad\times I_{q\leftarrow i}(z_1,x_T^2,\mu)\,
    I_{\bar q\leftarrow j}(z_2,x_T^2,\mu)
\end{aligned}
\end{equation}
with ordinary PDFs. The result reads
\begin{eqnarray}\label{fact1}
   \frac{d^3\sigma}{dM^2\,dq_T^2\,dy} 
   &=& \frac{4\pi\alpha^2}{3N_c M^2 s}\,\sum_q\,e_q^2\,\sum_{i=q,g} \sum_{j=\bar q,g} 
    \int_{\xi_1}^1\!\frac{dz_1}{z_1} \int_{\xi_2}^1\!\frac{dz_2}{z_2} \\
   &&\mbox{}\times \bigg[ C_{q\bar q\to ij}(z_1,z_2,q_T^2,M^2,\mu)\,
    \phi_{i/N_1}(\xi_1/z_1,\mu)\,\phi_{j/N_2}(\xi_2/z_2,\mu) 
    + (q,i\leftrightarrow\bar q,j) \bigg] \,. \nonumber
\end{eqnarray}
This formula, as well as relations (\ref{fact2}) and (\ref{fact3}) below, receive power corrections in the two small ratios $q_T^2/M^2$ and $\Lambda_{\rm QCD}^2/q_T^2$. This will not be indicated explicitly. 

Integrating this result over rapidity, with $|y|\le\ln(1/\tau)$, we obtain
\begin{equation}\label{fact2}
\begin{aligned}
   \frac{d^2\sigma}{dM^2\,dq_T^2} 
   &= \frac{4\pi\alpha^2}{3N_c M^2 s}\,\sum_q\,e_q^2\,\sum_{i=q,g} \sum_{j=\bar q,g}
    \hspace{3mm} \int\hspace{-4mm}\int\limits_{\hspace{-2.5mm}z_1 z_2\ge\tau}\!\!
    \frac{dz_1}{z_1}\,\frac{dz_2}{z_2} \\
   &\quad\times \bigg[ C_{q\bar q\to ij}(z_1,z_2,q_T^2,M^2,\mu)\,
    f\hspace{-1.8mm}f_{ij}\Big( \frac{\tau}{z_1 z_2},\mu \Big) 
    + (q,i\leftrightarrow\bar q,j) \bigg] \,,
\end{aligned}
\end{equation}
where the parton luminosities are defined as
\begin{equation}
   f\hspace{-1.8mm}f_{ij}(u,\mu)
   = \int_u^1\!\frac{dz}{z}\,\phi_{i/N_1}(z,\mu)\,\phi_{j/N_2}(u/z,\mu) \,.
\end{equation}
It will also be useful to study the total cross section defined with a cut $q_T\le Q_T$, which vetoes single jet emission. Neglecting the dependence of the variable $\tau$ in (\ref{taudef}) on $q_T^2$, which is a power-suppressed effect, we obtain from (\ref{fact2})
\begin{eqnarray}\label{fact3}
   \frac{d\sigma}{dM^2} \bigg|_{q_T\le Q_T}\!\! 
   &=& \frac{4\pi\alpha^2}{3N_c M^2 s}\,\sum_q\,e_q^2\,\sum_{i=q,g} \sum_{j=\bar q,g} 
    \hspace{6mm} \int\hspace{-7mm}\int\limits_{\hspace{-2.5mm}z_1 z_2\ge M^2/s}\!\!
    \frac{dz_1}{z_1}\,\frac{dz_2}{z_2} \\
   &&\times \bigg[ \int\limits_0^{{\rm min}(Q_T^2,\,z_1 z_2 s-M^2)}\!\!dq_T^2\,
    C_{q\bar q\to ij}(z_1,z_2,q_T^2,M^2,\mu)\,
    f\hspace{-1.8mm}f_{ij}\Big( \frac{M^2}{z_1 z_2 s},\mu \Big) 
    + (q,i\leftrightarrow\bar q,j) \bigg] \,. \nonumber
\end{eqnarray}

\section{\boldmath Calculation of the kernels $I_{q\leftarrow q}$ and $I_{q\leftarrow g}$}
\label{sec:oneloop}

We now perform a perturbative calculation of the relevant kernels $I_{i\leftarrow j}$ entering the factorization formula (\ref{Cdef}) at first non-trivial order in $\alpha_s$. Since we do not have explicit operator definitions of the refactorized transverse distribution functions $B_{i/N}$, we analyze instead the original functions ${\cal B}_{i/N}$ defined in (\ref{Bdef}), keeping in mind that only products of two such functions referring to different hadrons are well defined. If we write an operator-product expansion analogous to (\ref{OPE})
\begin{equation}
   {\cal B}_{i/N}(\xi,x_T^2,\mu) 
   = \sum_j \int_\xi^1\!\frac{dz}{z}\,{\cal I}_{i\leftarrow j}(z,x_T^2,\mu)\,
    \phi_{j/N}(\xi/z,\mu) + {\cal O}(\Lambda_{\rm QCD}^2\,x_T^2) \,,     
\end{equation}
it follows that the products of two ${\cal I}_{i\leftarrow j}$ functions are well defined and obey a factorization formula analogous to (\ref{refact}).

\subsection{One-loop results}
\label{sec:1loop}

\begin{figure}
\begin{center}
\begin{tabular}{ccccc}
\psfrag{hl}[b]{\footnotesize $hc$}
\psfrag{hr}[lb]{\footnotesize $hc$}
\psfrag{al}[b]{\footnotesize $\overline{hc}$}
\psfrag{ar}[lb]{\footnotesize $\overline{hc}$}
\psfrag{hac}{}
\psfrag{a}[rb]{\footnotesize $\beta$}
\psfrag{b}[lb]{\footnotesize $\alpha$}
\includegraphics[height=0.17\textwidth]{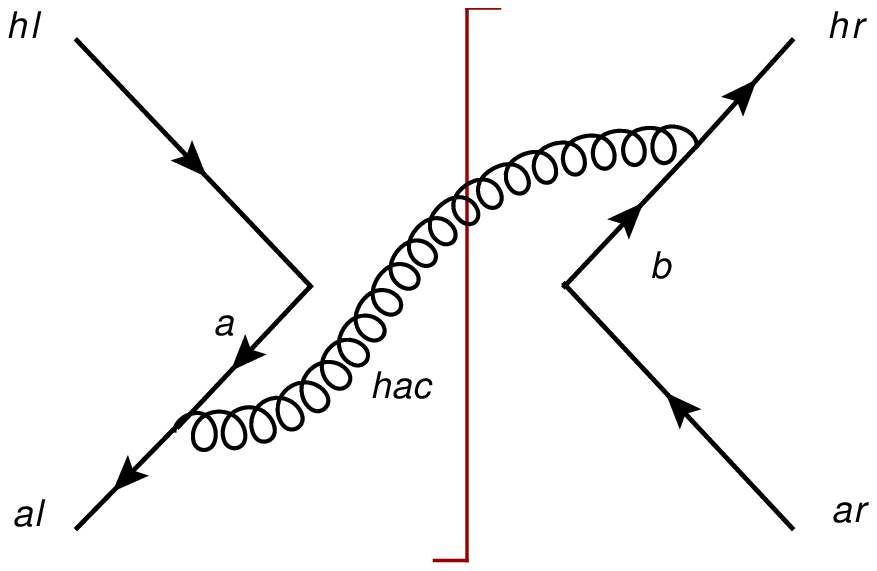} 
&\raisebox{1.3cm}{$\rightarrow$} & 
\psfrag{hl}[b]{\footnotesize $hc$}
\psfrag{hr}[lb]{\footnotesize $hc$}
\psfrag{al}[b]{\footnotesize $\overline{hc}$}
\psfrag{ar}[lb]{\footnotesize $\overline{hc}$}
\psfrag{a}[r]{\footnotesize $\beta$}
\psfrag{b}[lb]{\footnotesize $\alpha$}
\includegraphics[height=0.17\textwidth]{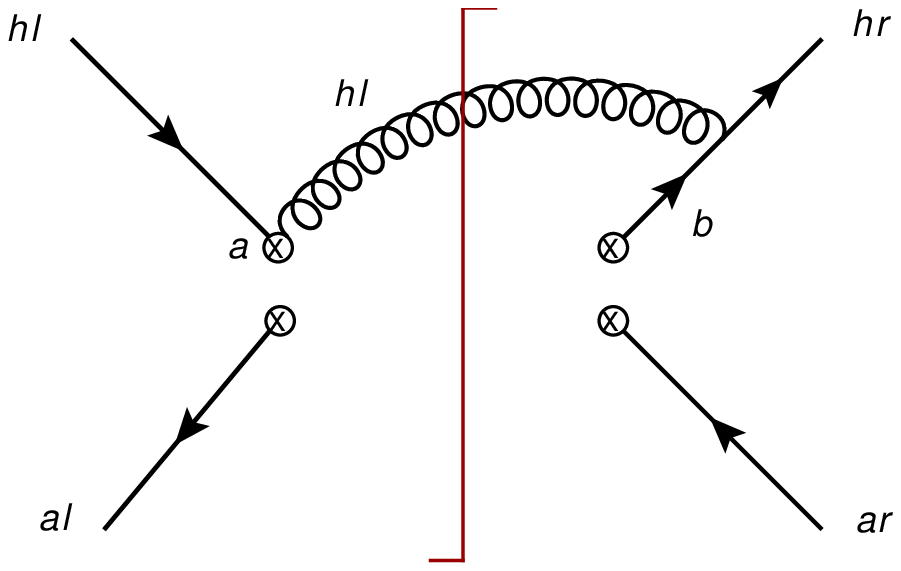} 
&\raisebox{1.3cm}{$+$} &  
\psfrag{hl}[b]{\footnotesize $hc$}
\psfrag{hr}[lb]{\footnotesize $hc$}
\psfrag{al}[b]{\footnotesize $\overline{hc}$}
\psfrag{ar}[lb]{\footnotesize $\overline{hc}$}
\psfrag{a}[rb]{\footnotesize $\beta$}
\psfrag{b}[lb]{\footnotesize $\,\alpha$}
\includegraphics[height=0.17\textwidth]{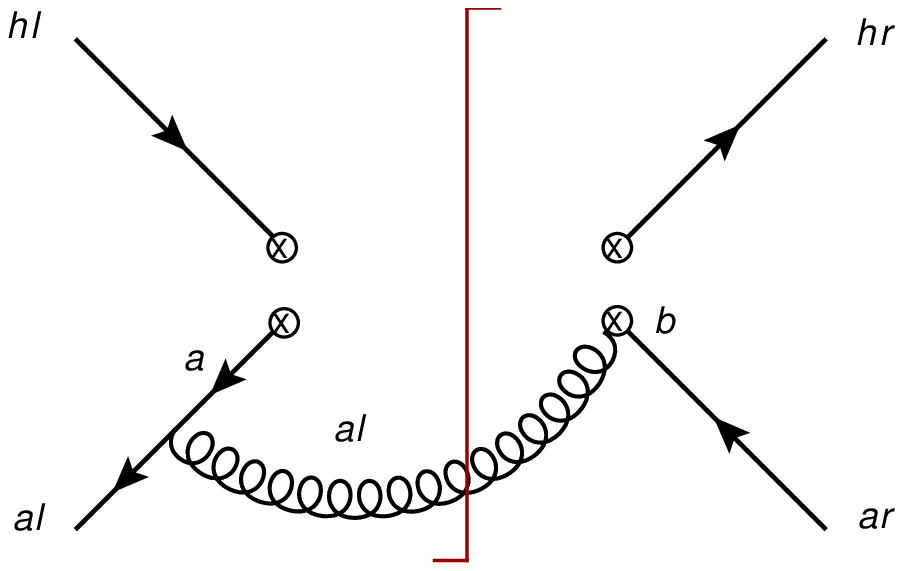}  
\end{tabular}
\vspace{-0.5cm}
\end{center}
\caption{\label{fig:match}
Matching of an analytically-regularized QCD graph onto SCET diagrams.}
\end{figure}

Perturbative expansions for the kernels ${\cal I}_{i\leftarrow j}$ can be derived from a matching calculation, in which the matrix elements in (\ref{phidef}) and (\ref{Bdef}) are evaluated using external parton states carrying a fixed fraction of the nucleon momentum $p$. The tree-level result is obviously given by
\begin{equation}\label{Itree}
   {\cal I}_{i\leftarrow j}(z,x_T^2,\mu) 
   = \delta(1-z)\,\delta_{ij} + {\cal O}(\alpha_s) \,.
\end{equation}
However, when trying to evaluate the one-loop corrections, one finds that they are ill-defined in dimensional regularization due to light-cone singularities. To give meaning to the corresponding loop integrals requires introducing additional regulators. The simplest possibility is to employ analytic regularization, as is common in the context of asymptotic expansions \cite{Smirnov:1993ta,Smirnov:2002pj}. In the context of SCET this method has been used in \cite{Chiu:2007yn,Beneke:2003pa}. One starts by reconsidering the QCD diagrams contributing to the process and raises all propagators through which the external hard-collinear momentum $p$ flows to a fractional power,
\begin{equation}\label{analytreg}
   \frac{1}{-(p-k)^2-i\varepsilon} 
   \to \frac{\nu_1^{2\alpha}}{\left[-(p-k)^2-i\varepsilon\right]^{1+\alpha}} \,,
\end{equation}
and similarly for the anti-hard-collinear propagators, but with a different regulator $\beta$ and an associated scale $\nu_2$. For QCD diagrams, such as the first graph in Figure~\ref{fig:match}, the modification is trivial in the sense that the limits $\alpha\to 0$ and $\beta\to 0$ are smooth as long as the dimensional regulator $d=4-2\epsilon$ is kept in place. However, with analytic regulators the contributions of the different momentum regions are now well-defined individually and one can check which regions give non-vanishing contributions to the expansion of the loop integrals. One finds that only the hard, hard-collinear, and anti-hard-collinear regions contribute. If the diagrams are evaluated off-shell, then also a soft contribution arises in the individual diagrams, but one easily verifies that the contribution of the semi-hard mode vanishes. Since such contributions were considered in the literature, we now explicitly show that they are absent in analytical regularization. To do so, one assumes that the gluon momentum $k$ in the QCD diagram in Figure~\ref{fig:match} scales as $M(\lambda,\lambda,\lambda)$ and then expands the diagram  in powers of $\lambda$. At leading power, the relevant phase-space integral becomes
\begin{equation}\label{Ish}
   \int\!d^dk\,\frac{1}{\left( n\cdot k - i\epsilon \right)^{1+\alpha}}\,
   \frac{1}{\left( \bar{n}\cdot k - i\epsilon \right)^{1+\beta}}\,\delta(k^2)\, 
   \theta(k^0)\,e^{ip\cdot x - ik_\perp\cdot x_\perp} \,.
\end{equation}
After the expansion, the integrand involves the usual eikonal propagators characteristic for soft emissions, which are raised to fractional powers because of the analytic regulators. It is important that not only the propagator denominators, but also the Fourier exponent is multipole expanded:
\begin{equation}
   (p-k)\cdot x = p\cdot x - k_\perp\cdot x_\perp + {\cal O}(\lambda) \,,
\end{equation}
which follows from the scaling $x\sim(1,1,\lambda^{-1})$ derived in Section \ref{sec: factorization}. Performing the integration over the light-cone components of the gluon momentum, one finds that the integral (\ref{Ish}) is scaleless and vanishes. The same argument applies to multi-loop integrals involving semi-hard propagators.

We note that the multipole expansion was not performed in \cite{Mantry:2009qz}, which explains why similar integrals were found to be non-vanishing in this reference. However, the expansion is a crucial ingredient to achieve scale separation for effective theories in dimensional regularization. It is equally important in the strategy of region technique \cite{Smirnov:1993ta,Smirnov:2002pj}. Without performing the expansion integrals pick up contributions from several regions, and care needs to be taken to avoid double counting.

Having shown that the soft and semi-hard regions do not contribute, let us now turn to the hard-collinear contributions. Since the original diagrams are well defined without analytical regulators and are obtained by adding up the contributions from the different regions, we are guaranteed that the limits $\alpha\to 0$ and $\beta\to 0$ can be taken in the sum of all diagrams and that the final result is independent of the regularization scheme. Individually, however, the diagrams in each sector involve divergences in the analytical regulators. If the momentum $k$ in (\ref{analytreg}) is hard-collinear, as in the first SCET diagram in Figure~\ref{fig:match}, the $\alpha$ dependence in the effective theory takes the same form as in QCD. If, on the other hand, the momentum $k$ is anti-hard-collinear, then the propagator is far off-shell and in SCET is represented by a Wilson line, as shown in the second diagram in Figure~\ref{fig:match}. Using the replacement rule (\ref{analytreg}) and performing the appropriate expansions, we find that the Feynman rule for a gluon emission from the anti-hard-collinear Wilson line $W_{\overline{hc}}$ in the current operator (\ref{eq:current}) gets replaced by
\begin{equation}\label{wilsonreg}
   \frac{n^\mu}{n\cdot k-i\varepsilon} 
   \to \frac{\nu_1^{2\alpha}\,\,n^\mu\,\bar n\cdot p}%
            {\left(n\cdot k\,\bar n\cdot p-i\varepsilon\right)^{1+\alpha}} \,.
\end{equation}
Note that, as mentioned earlier in Section~\ref{sec:collinearanomaly}, the regularized Feynman rule for the anti-hard-collinear Wilson line is no longer invariant under the rescaling transformation $p\to\lambda p$. As seen in Figure~\ref{fig:match}, the regulator $\alpha$ plays a double role: it regularizes the fermion propagators in hard-collinear diagrams and the Wilson lines in anti-hard-collinear diagrams. Both classes of diagrams develop singularities in the limit $\beta\to 0$ followed by $\alpha\to 0$ or vice versa, which cancel in the sum of the results from both sectors.

\begin{figure}
\begin{center}
\begin{tabular}{cccc}
\includegraphics[height=0.142\textwidth]{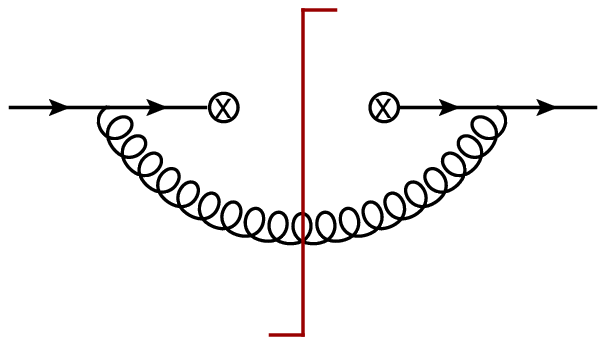} & 
\includegraphics[height=0.142\textwidth]{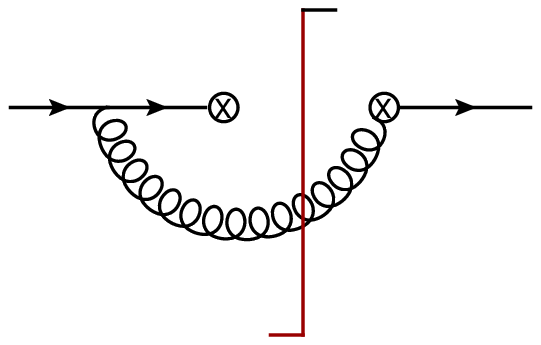} & 
\includegraphics[height=0.142\textwidth]{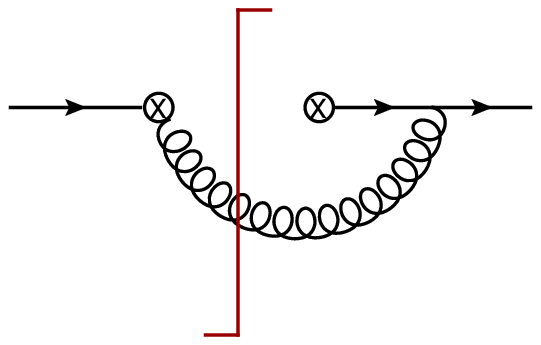} & 
\includegraphics[height=0.142\textwidth]{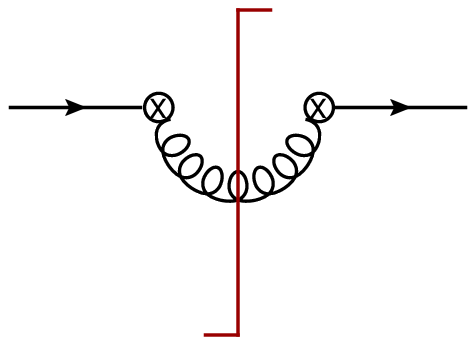} \\[0.1cm]
\end{tabular}
\includegraphics[height=0.142\textwidth]{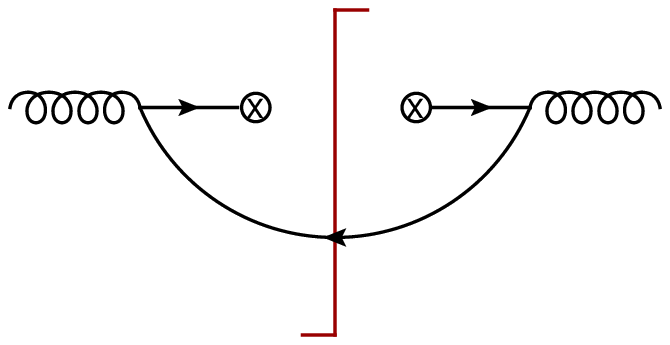}
\vspace{-0.5cm}
\end{center}
\caption{\label{fig:graphs}
One-loop diagrams contributing to the matching coefficients ${\cal I}_{q\leftarrow q}$ (top row) and ${\cal I}_{q\leftarrow g}$ (bottom row). The vertical lines indicate cut propagators.}
\end{figure}

With the regularization in place, let us now turn to the evaluation of the one-loop corrections to the kernels ${\cal I}_{q\leftarrow q}={\cal I}_{\bar q\leftarrow\bar q}$. The relevant effective theory diagrams are shown in the first row of Figure~\ref{fig:graphs}. There is no need to consider diagrams with external-leg corrections on only one side of the cut, because these give identical contributions to ${\cal B}_{i/N}$ and $\phi_{i/N}$ and thus do not change the tree-level result (\ref{Itree}). Working in Feynman gauge, we find that the contribution of the first diagram is finite in the limit where the analytic regulators are sent to zero as long as $\epsilon$ is finite. A divergence arises only in the expansion in $\epsilon$, after which we obtain
 \begin{equation}\label{Lperpdef}
   {\cal I}_{q\leftarrow q}^{a}(z,x_T^2,\mu) 
   = - \frac{C_F\alpha_s}{2\pi}\,(1-z) \left( \frac{1}{\epsilon} + L_\perp - 1 
    \right) , \qquad
   L_\perp = \ln\frac{x_T^2\mu^2}{4e^{-2\gamma_E}} \,,
\end{equation}
up to ${\cal O}(\epsilon)$ terms. As before, $\alpha_s\equiv\alpha_s(\mu)$ always refers to the running coupling evaluated at the scale $\mu$, unless indicated otherwise. Moreover, $\mu$ denotes the renormalization scale defined in the $\overline{\rm MS}$ scheme. The fourth diagram gives a vanishing result, ${\cal I}_{q\leftarrow q}^{d}=0$, but the remaining two diagrams in Figure~\ref{fig:graphs} are non-zero and only well-defined with analytic regulators. Both give the same result, and for their sum we obtain
\begin{equation}
   {\cal I}_{q\leftarrow q}^{b+c}(z,x_T^2,\mu) 
   = \frac{C_F\alpha_s}{2\pi}\,e^{\epsilon\gamma_E}
    \left( \frac{\mu^2}{\nu_1^2} \right)^{-\alpha} 
    \left( \frac{q^2}{\nu_2^2} \right)^{-\beta} 
    \frac{2z}{(1-z)^{1-\alpha+\beta}}\,
    \frac{\Gamma(-\epsilon-\alpha)}{\Gamma(1+\alpha)}
    \left( \frac{x_T^2\mu^2}{4} \right)^{\epsilon+\alpha} .
\end{equation}
Like in full QCD, the analytic regulators must be taken to zero before taking the limit $\epsilon\to 0$. The result depends on the order in which the limits $\alpha\to 0$ and $\beta\to 0$ are performed. Expanding first in $\beta$ and then in $\alpha$, the light-cone singularities are regulated by the $\alpha$ parameter, and we find for the sum of all four one-loop diagrams
\begin{eqnarray}\label{Iqqbarealpha}
   {\cal I}_{q\leftarrow q}(z,x_T^2,\mu) \Big|_{\alpha~{\rm reg.}}
   &=& - \frac{C_F\alpha_s}{2\pi}\,\bigg\{ \!
    \left( \frac{1}{\epsilon} + L_\perp \right)
    \left[ \left( \frac{2}{\alpha} - 2\ln\frac{\mu^2}{\nu_1^2} \right) \delta(1-z) 
    + \frac{1+z^2}{(1-z)_+} \right] \nonumber\\
   &&\hspace{17mm}\mbox{}+ \delta(1-z) \left( - \frac{2}{\epsilon^2} + L_\perp^2 
    + \frac{\pi^2}{6} \right) - (1-z) \bigg\} \,. 
\end{eqnarray}
If the expansions are performed in the opposite order, then $\beta$ acts as the analytic regulator, and we obtain
\begin{equation}\label{Iqqbarebeta}
   {\cal I}_{q\leftarrow q}(z,x_T^2,\mu) \Big|_{\beta~{\rm reg.}}
   = - \frac{C_F\alpha_s}{2\pi}\,\bigg\{ \! \left( \frac{1}{\epsilon} + L_\perp \right)
    \left[ \left( - \frac{2}{\beta} + 2\ln\frac{q^2}{\nu_2^2} \right) \delta(1-z) 
    + \frac{1+z^2}{(1-z)_+} \right] - (1-z) \bigg\} \,. 
\end{equation}

The above results refer to the kernel associated with hard-collinear partons, which propagate along the $n$ direction. Let us now consider what happens when we calculate the corresponding kernel for anti-hard-collinear fields. In this case we get the same answer but with $\alpha,\nu_1$ and $\beta,\nu_2$ interchanged. We then find that in the product of a hard-collinear and an anti-hard-collinear kernel function the analytic regulators disappear, no matter in which order the limits $\alpha\to 0$ and $\beta\to 0$ are taken. This product is thus regulator independent and well defined in dimensional regularization. After $\overline{\rm MS}$ subtractions, we obtain
\begin{equation}\label{Ifinal}
\begin{aligned}
   &\left[ {\cal I}_{q\leftarrow q}(z_1,x_T^2,\mu)\,
    {\cal I}_{\bar q\leftarrow\bar q}(z_2,x_T^2,\mu) \right]_{q^2} \\
   &= \delta(1-z_1)\,\delta(1-z_2) \left[ 1 - \frac{C_F\alpha_s}{2\pi}
    \left( 2L_\perp \ln\frac{q^2}{\mu^2} + L_\perp^2 - 3L_\perp + \frac{\pi^2}{6}
    \right) \right] \\
   &\quad\mbox{}- \frac{C_F\alpha_s}{2\pi}\,\bigg\{ \delta(1-z_1) 
    \left[ L_\perp \left( \frac{1+z_2^2}{1-z_2} \right)_+ - (1-z_2) \right] 
    + (z_1\leftrightarrow z_2) \bigg\} + {\cal O}(\alpha_s^2) \,. 
\end{aligned}
\end{equation}

Next we calculate the kernel ${\cal I}_{q\leftarrow g}$ at one-loop order. This function vanishes at tree level, and at one-loop order it follows from the evaluation of the diagram shown in the second row of Figure~\ref{fig:graphs}. There is no need for analytic regularization in this case, and after $\overline{\rm MS}$ subtractions we find
\begin{equation}\label{Ifinal2}
   {\cal I}_{q\leftarrow g}(z,x_T^2,\mu) 
   = - \frac{T_F\alpha_s}{2\pi}\,\Big\{ L_\perp 
    \big[ z^2 + (1-z)^2 \big] - 2z(1-z) \Big\} + {\cal O}(\alpha_s^2) \,.
\end{equation}

From (\ref{Ifinal}) and (\ref{Ifinal2}) we can extract the one-loop expressions for the renormalized function $F_{q\bar q}$ and the renormalized kernels $I_{i\leftarrow j}$ relevant for Drell-Yan production. We obtain
\begin{equation}\label{F1loop}
   F_{q\bar q}(L_\perp,\alpha_s)
   = \frac{C_F\alpha_s}{\pi}\,L_\perp + {\cal O}(\alpha_s^2) \,, \\
\end{equation}
and
\begin{equation}\label{Iresults}
\begin{aligned}
   I_{q\leftarrow q}(z,L_\perp,\alpha_s)
   &= \delta(1-z) \left[ 1 + \frac{C_F\alpha_s}{4\pi}
    \left( L_\perp^2 + 3 L_\perp - \frac{\pi^2}{6} \right) \right] \\
   &\quad\mbox{}
   - \frac{C_F\alpha_s}{2\pi}\,
    \Big[ L_\perp P_{q\leftarrow q}(z) - (1-z) \Big] + {\cal O}(\alpha_s^2) \,, \\
   I_{q\leftarrow g}(z,L_\perp,\alpha_s)
   &= - \frac{T_F\alpha_s}{2\pi}\,\Big[ L_\perp P_{q\leftarrow g}(z) - 2z(1-z) \Big]
    + {\cal O}(\alpha_s^2) \,,     
\end{aligned}
\end{equation}
with $L_\perp$ as defined in (\ref{Lperpdef}). The kernel $I_{\bar q\to\bar q}$ is given by the same expression as $I_{q\to q}$, and $I_{\bar q\to g}$ has the same form as $I_{q\to g}$. Note that, with a slight abuse of notation, we have changed the arguments $x_T^2$ and $\mu$ in $F_{q\bar q}$ and the kernel functions to $L_\perp$ and $\alpha_s$, as this will be more convenient from now on. In the above expressions
\begin{equation}\label{AP1loop}
   P_{q\leftarrow q}(z) = \left( \frac{1+z^2}{1-z} \right)_+ \,, \qquad
   P_{q\leftarrow g}(z) = z^2 + (1-z)^2
\end{equation}
are the one-loop Altarelli-Parisi splitting functions, defined as  
\begin{equation}
   {\cal P}_{q\leftarrow q}(x,\mu) 
   = \frac{C_F\alpha_s}{\pi}\,P_{q\leftarrow q}(x) + {\cal O}(\alpha_s^2) \,, 
    \qquad
   {\cal P}_{q\leftarrow g}(x,\mu) 
   = \frac{T_F\alpha_s}{\pi}\,P_{q\leftarrow g}(x) + {\cal O}(\alpha_s^2) \,.
\end{equation}
It is straightforward to check that our one-loop results (\ref{Ifinal}) and (\ref{Ifinal2}) satisfy the general evolution equations (\ref{Ievol}).

\subsection{All-order dependence on the hard momentum transfer}
\label{sec:refact}

The appearance of a logarithm of the large momentum transfer $q^2$ in the matching condition (\ref{Ifinal}) appears strange at first sight, since it arises from the evaluation of hard-collinear and anti-hard-collinear loop graphs in the effective theory, in which these two sectors are decoupled from each other at the Lagrangian level. Naively, we would thus expect a dependence on the scale $x_T$ only. For $\mu$ of order a typical hard-collinear scale the resulting logarithm is parametrically large, so that $\alpha_s\ln(q^2/\mu^2)\sim 1$ in RG power counting. The question then arises if higher powers of such logarithms appear in higher orders of perturbation theory, and if this is the case, how these logarithms can be resummed to all orders. 

A simpler example for the occurrence of the collinear anomaly is the Sudakov form factor of a massive vector boson, which in SCET has been discussed in \cite{Chiu:2007yn,Chiu:2007dg}. The similarity with our case arises because the $\delta$-function requiring that the square of the total transverse momentum $k_{\perp {\rm tot}}$ of all cut propagators in the discontinuity of the forward-scattering amplitude be equal to $q_\perp^2$ can be rewritten as the discontinuity of a ``propagator'' $1/(k_{\perp {\rm tot}}^2+q_T^2)$, which indeed looks similar to a massive propagator. Using arguments along the lines described in \cite{Chiu:2007dg}, we will now argue that the {\em logarithm\/} of the product of two transverse PDFs ${\cal B}_{i/N_1}\,{\cal B}_{j/N_2}$, or of two kernels ${\cal I}_{i\leftarrow k}\,{\cal I}_{j\leftarrow l}$, is linear in $\ln(q^2/\mu^2)$ with a universal coefficient $-F_{ij}(x_T^2,\mu)$ that is independent of $z_1$, $z_2$ and $k$, $l$, as shown in (\ref{refact}). This result implies that the large logarithms exponentiate in $x_T$ space.

Because of the formal decoupling of the hard-collinear and anti-hard-collinear sectors in the SCET Lagrangian, the product ${\cal B}_{i/N_1}\,{\cal B}_{j/N_2}$ is a product of the sum of all hard-collinear graphs times the sum of all anti-hard-collinear graphs, each defined by means of analytic regulators. Let us denote the logarithms of the sums of all (anti-)hard-collinear graphs by
\begin{equation}\label{HH}
\begin{aligned}
   \ln{\cal B}_{i/N_1} = \ln(\sum \mbox{$hc$ graphs}) 
   &\equiv {\cal H}_i\bigg(\xi_1,L_\perp,\alpha_s(\mu),\ln\frac{\mu^2}{\nu_1^2},
    \ln\frac{q^2}{\nu_2^2}\bigg) \,, \\
   \ln{\cal B}_{j/N_2} = \ln(\sum \mbox{$\overline{hc}$ graphs}) 
   &\equiv \overline{\cal H}_j\bigg(\xi_2,L_\perp,\alpha_s(\mu),\ln\frac{q^2}{\nu_1^2},
    \ln\frac{\mu^2}{\nu_2^2}\bigg) \,.
\end{aligned}
\end{equation}
The need for analytic regularization introduces logarithmic dependence on the auxiliary scales $\nu_1$ and $\nu_2$ associated with the $\alpha$ and $\beta$ regulators, see Section~\ref{sec:1loop}. The precise form of the results for ${\cal H}_i$ and $\overline{\cal H}_j$ (but not for their sum) depends on the order in which the limits $\alpha\to 0$ and $\beta\to 0$ are taken. All that matters is that the same prescription is used throughout the calculation. In the hard-collinear sector the logarithmic dependence on $\nu_1$ can for dimensional reasons only appear in the form of $\ln(\mu^2/\nu_1^2)$, since no other momentum scale is introduced by the $\alpha$ regulator. In the anti-hard-collinear sector, on the other hand, the induced logarithmic dependence on $\nu_1$ is accompanied by $n\cdot\bar p$. By Lorentz invariance, a logarithm of the ``foreign'' momentum component $n\cdot\bar p$ can only appear in the form of $n\cdot\bar p\,\bar n\cdot p=q^2$, so that we encounter $\ln(q^2/\nu_1^2)$. For the $\beta$ regulator with associated scale $\nu_2$, the role of the two sectors is reversed. This explains the dependencies on the two regulator scales shown in (\ref{HH}). 

The requirement that the product of the two transverse PDFs must be independent of the two regulator scales implies that
\begin{equation}
   \frac{d}{d\ln\nu_1} \left( {\cal H}_i + \overline{\cal H}_j \right) = 0 
   = \frac{d}{d\ln\nu_2} \left( {\cal H}_i + \overline{\cal H}_j \right) .
\end{equation}
These conditions enforce that ${\cal H}_i$ and $\overline{\cal H}_j$ are linear in their last two arguments with coefficients that are independent of $\xi_1$ and $\xi_2$, and hence
we are free to write
\begin{equation}
   \ln{\cal B}_{i/N_1}\,{\cal B}_{j/N_2} 
   = H_i\big(\xi_2,L_\perp,\alpha_s(\mu)\big) 
    + \overline{H}_j\big(\xi_2,L_\perp,\alpha_s(\mu)\big)
    - F_{ij}\big(L_\perp,\alpha_s(\mu)\big) 
    \left( \ln\frac{q^2}{\mu^2} + L_\perp \right) .
\end{equation}
We have used the freedom that the decomposition is unique up to a function of $L_\perp$ to make the coefficient of $F_{q\bar q}$ scale independent. With the identification $H_i=\ln B_{i/N_1}$ and $\overline{H}_j=\ln B_{j/N_2}$ this proves relation (\ref{refact}), where we have considered the special case $i=q$, $j=\bar q$.

Let us now collect what can be said about the function $F_{q\bar q}$ based on general principles. Generalizing our one-loop result to higher orders, we can write the perturbative expansion of $F_{q\bar q}$ in the form
\begin{equation}\label{Fsum}
   F_{q\bar q}(L_\perp,\alpha_s) = \sum_{n=1}^\infty\,d_n^q(L_\perp) 
    \left( \frac{\alpha_s}{4\pi} \right)^n ,
\end{equation}
where $d_1^q(L_\perp)=4C_F L_\perp$. The first evolution equation in (\ref{Bevol}) then implies the recursion relation
\begin{equation}
   {d_n^q}'(L_\perp)
   = \Gamma_{n-1}^F + \sum_{m=1}^{n-1}\,m\,\beta_{n-1-m}\,d_m^q(L_\perp) \,,
    \qquad n\ge 1 \,,
\end{equation}
where the prime denotes a derivative with respect to $L_\perp$, and as usual we have expanded the cusp anomalous dimension and $\beta(\alpha_s)=\mu\,d\alpha_s/d\mu$ as
\begin{equation}
   \Gamma_{\rm cusp}^F(\alpha_s) 
   = \sum_{n=1}^\infty\,\Gamma_{n-1}^F \left( \frac{\alpha_s}{4\pi} \right)^n , 
    \qquad
   \beta(\alpha_s) = -2\alpha_s \sum_{n=1}^\infty\,\beta_{n-1}
    \left( \frac{\alpha_s}{4\pi} \right)^n .
\end{equation}
For the first two expansion coefficients, we obtain
\begin{equation}\label{c1c2}
   d_1^q(L_\perp) = \Gamma_0^F L_\perp + d_1^q \,,
   \qquad
   d_2^q(L_\perp) = \frac{\Gamma_0^F\beta_0}{2}\,L_\perp^2 + \Gamma_1^F L_\perp + d_2^q \,,
\end{equation}
where $d_n^q\equiv d_n^q(0)$ with $d_1^q=0$. The expansion of the corresponding function $F_{gg}$ for Higgs production can be written as in (\ref{Fsum}) but with coefficients $d_n^g$, which obey analogous equations in which $\Gamma_{\rm cusp}^F$ is replaced by $\Gamma_{\rm cusp}^A$. We will later discuss how the two-loop coefficients $d_2^{q,g}$ can be extracted from existing calculations of higher-order corrections to Drell-Yan and Higgs production cross sections derived in fixed-order perturbation theory \cite{Davies:1984sp,deFlorian:2001zd}. The result is
\begin{equation}\label{d2resu}
   \frac{d_2^q}{C_F} =\frac{d_2^g}{C_A} 
   = C_A \left( \frac{808}{27} - 28\zeta_3 \right) - \frac{224}{27}\,T_F n_f \,.
\end{equation}
These coefficients contain only maximally non-abelian color structures. This leads us to conjecture that also in higher orders they are constrained by the non-abelian exponentiation theorem \cite{Gatheral:1983cz,Frenkel:1984pz}, as is the case for the cusp anomalous dimension. This would imply that the Casimir scaling relation $d_n^q/C_F=d_n^g/C_A$ continues to hold at least to three-loop order. Since the cusp anomalous dimension obeys the same relation, Casimir scaling to three-loop order holds for the entire $F_{q\bar q}$ and $F_{gg}$ functions, as shown in (\ref{Casimir}). Note that there are arguments indicating that for the cusp anomalous dimension Casimir scaling should hold at four loops and perhaps even to all orders of perturbation theory \cite{Becher:2009qa}.

\section{Resummation and Fourier transformation}
\label{sec:resum}

In the differential cross section (\ref{Bfact}) and the expression for the hard-scattering kernels $C_{q\bar q\to ij}$ in (\ref{Cdef}) the dependence on the scales $M^2$ and $x_T^2$ is factorized explicitly. However, for any given choice of the renormalization scale $\mu$ these expressions contain large logarithms, which need to be resummed to all orders in perturbation theory. This is accomplished by solving RG equations for the various component functions. We will now discuss these solutions in detail. We will then derive a compact, all-order formula for the hard-scattering kernels $C_{q\bar q\to ij}$ in $q_T$ space, which is free of large logarithms.

In practice, the easiest way to perform the resummation of large logarithms is to choose $\mu$ of order a typical hard-collinear scale, i.e.\ $\mu\sim q_T$ or $\mu\sim x_T^{-1}$, and then evolve the hard function and the PDFs from appropriate initial scales to the scale $\mu$. The PDF evolution is standard and can be taken from any package that generates parton distributions. The evolution of the hard function reads \cite{Becher:2006mr}
\begin{equation}\label{CVsol}
   C_V(-M^2,\mu) 
   = \exp\left[ 2S(\mu_h,\mu) - 2a_{\gamma^q}(\mu_h,\mu) \right]
    \left( \frac{-M^2}{\mu_h^2} \right)^{-a_\Gamma(\mu_h,\mu)}\,C_V(-M^2,\mu_h) \,,
\end{equation}
where $\mu_h^2\sim-M^2$ is a hard matching scale, at which the value of $C_V$ is calculated using fixed-order perturbation theory. At one-loop order \cite{Manohar:2003vb,Becher:2006mr}
\begin{equation}
   C_V(-M^2,\mu_h) 
   = 1 + \frac{C_F\alpha_s(\mu_h)}{4\pi}
    \left( - L^2 + 3L - 8 + \frac{\pi^2}{6} \right) + \dots \,,
\end{equation}
where $L=\ln(-M^2/\mu_h^2)$. The two-loop correction can be found in \cite{Becher:2006mr}. The advantages of using a time-like scale choice ($\mu_h^2<0$) for time-like processes such as Drell-Yan production were emphasized in \cite{Ahrens:2008qu,Ahrens:2008nc,Magnea:1990zb}. The Sudakov exponent $S$ and the exponents $a_n$ are given by \cite{Neubert:2004dd}
\begin{equation}\label{RGEsols}
   S(\nu,\mu) 
   = - \int_\nu^\mu\!\frac{d\bar\mu}{\bar\mu}\,\ln\frac{\bar\mu}{\nu}\,
    \Gamma_{\rm cusp}^F\big(\alpha_s(\bar\mu)\big) \,, \qquad   
   a_\Gamma(\nu,\mu) 
   = - \int_\nu^\mu\!\frac{d\bar\mu}{\bar\mu}\,
    \Gamma_{\rm cusp}^F\big(\alpha_s(\bar\mu)\big) \,,    
\end{equation}
and similarly for the function $a_{\gamma^q}$. Note that for gluon-initiated processes the functions $S$ and $a_\Gamma$ would (at least up to three-loop order) simply be rescaled by $C_A/C_F$. The solutions to the evolution equations (\ref{Bevol}) for the functions $F_{q\bar q}$ and $B_{q/N}$ can be expressed in terms of the same objects. They are given in Appendix~\ref{app:a}, where we also collect the perturbative expansions of the anomalous dimensions and the resulting expressions for the evolution functions valid at next-to-leading order in RG-improved perturbation theory.

Besides the large logarithms resummed in the evolution of the hard function $|C_V|^2$, there are large logarithms arising from the collinear anomaly, which are contained in the first factor on the right-hand side of (\ref{refact}). In order to discuss the resummation of these logarithms in a systematic manner, we first rewrite
\begin{equation}
   \left( \frac{x_T^2 M^2}{4e^{-2\gamma_E}} \right)^{-F_{q\bar q}(x_T^2,\mu)}
   = \left( \frac{M^2}{\mu^2} \right)^{-F_{q\bar q}(L_\perp,\alpha_s)}
    e^{-L_\perp F_{q\bar q}(L_\perp,\alpha_s)} \,.
\end{equation}
While this formula is exact, it does not yet provide a convenient basis for a perturbative analysis of the resummed kernels in RG-improved perturbation theory. Counting as usual the large logarithm $\ln(M^2/\mu^2)$ like $1/\alpha_s$, we see that the first term in the perturbative series (\ref{Fsum}) for the exponent $F_{q\bar q}$ needs to be evaluated in exponentiated form, while from two-loop order on a perturbative expansion can be applied. We thus split
\begin{equation}
   F_{q\bar q}(L_\perp,\alpha_s) 
   = \frac{\Gamma_0^F\alpha_s}{4\pi}\,L_\perp + \sum_{n=2}^\infty\,d_n^q(L_\perp) 
    \left( \frac{\alpha_s}{4\pi} \right)^n 
   \equiv \frac{\Gamma_0^F\alpha_s}{4\pi} 
    \left[ L_\perp + f_{q\bar q}(L_\perp,\alpha_s) \right] ,
\end{equation}
where the function $f_{q\bar q}$ starts at ${\cal O}(\alpha_s)$ but only receives contributions of two-loop order and higher. We now have
\begin{equation}\label{vari2}
  \left( \frac{M^2}{\mu^2} \right)^{-F_{q\bar q}(L_\perp,\alpha_s)}
  = \left( \frac{x_T^2\mu^2}{4e^{-2\gamma_E}} \right)^{-\eta_F(M^2,\mu)}
   e^{-\eta_F(M^2,\mu)\,f_{q\bar q}(L_\perp,\alpha_s)} \,,
\end{equation}
where
\begin{equation}\label{etaQ}
   \eta_F(M^2,\mu) = \frac{C_F\alpha_s}{\pi}\,\ln\frac{M^2}{\mu^2}
\end{equation}
counts as an ${\cal O}(1)$ quantity as long as $\mu^2\ll M^2$. 

Our last task is to perform the Fourier transformation (\ref{Cdef}) from $x_T$ space to transverse-momentum space. Since all our functions depend on $x_T$ either via a power, like in (\ref{vari2}), or via the logarithm $L_\perp$ defined in (\ref{Lperpdef}), it suffices to consider the relation 
\begin{equation}\label{Fourier}
   \frac{1}{4\pi} \int\!d^2x_\perp\,e^{-iq_\perp\cdot x_\perp}\,L_\perp^n
    \left( \frac{x_T^2\mu^2}{4e^{-2\gamma_E}} \right)^{-\eta}
   = (-\partial_\eta)^n\,\frac{1}{q_T^2} \left( \frac{q_T^2}{\mu^2} \right)^\eta 
    \frac{\Gamma(1-\eta)}{e^{2\eta\gamma_E}\,\Gamma(\eta)} \,.
\end{equation}
While $q_T$ serves as an infrared regulator, the integral converges in the ultraviolet, for $x_T\to 0$, only as long as $\eta<1$. In our formulae below we will always assume that the condition $\eta_F(M^2,\mu)<1$ is fulfilled.\footnote{The singularity at $\eta=1$ was noted a long time ago and referred to as a ``geometrical singularity'' \cite{Frixione:1998dw}. We will show in Section~\ref{sec:asymp} that it disappears when a class of factorially divergent higher-order terms is resummed.}  
For $M=M_Z$ or $M_W$, for example, this implies that $\mu$ should be larger than about 2\,GeV. Note that the higher-derivative terms in (\ref{Fourier}) are accompanied by powers of $1/(1-\eta)$, so that for $\eta$ very close to 1 a reorganization of the perturbative series becomes necessary. This will be discussed in detail in a forthcoming article \cite{inprep}.

Using the above result, we obtain a closed-form expression for the resummed hard-scattering kernels, which reads  
\begin{equation}\label{master}
\begin{aligned}
   C_{q\bar q\to ij}(z_1,z_2,q_T^2,M^2,\mu)
   &= \left| C_V(-M^2,\mu) \right|^2 I_{q\leftarrow i}(z_1,-\partial_\eta,\alpha_s)\,
    I_{\bar q\leftarrow j}(z_2,-\partial_\eta,\alpha_s) \\
   &\quad\times E_{q\bar q}(-\partial_\eta,\alpha_s,\eta_F)\,
    \frac{1}{q_T^2} \left( \frac{q_T^2}{\mu^2} \right)^\eta
    \frac{\Gamma(1-\eta)}{e^{2\eta\gamma_E}\,\Gamma(\eta)} \bigg|_{\eta=\eta_F} , 
\end{aligned}
\end{equation}
where $\eta_F\equiv\eta_F(M^2,\mu)$, and the arguments of the $I_{i\leftarrow j}(z,L_\perp,\alpha_s)$ functions are those shown in (\ref{Iresults}). It is understood that for $|C_V|^2$ one uses the resummed expression in (\ref{CVsol}). All remaining quantities have perturbative expansions in powers of $\alpha_s=\alpha_s(\mu)$ free of large logarithms. In writing the above result we have introduced the function
\begin{equation}\label{beauty}
\begin{aligned}
   E_{q\bar q}(L_\perp,\alpha_s,\eta) 
   &= \exp\Big[ -L_\perp F_{q\bar q}(L_\perp,\alpha_s) 
    -\eta\,f_{q\bar q}(L_\perp,\alpha_s) \Big] \\
   &= 1 - \frac{\alpha_s}{4\pi} \left[ \Gamma_0^F\,L_\perp^2
    + \eta \left( \frac{\beta_0}{2}\,L_\perp^2
    + \frac{\Gamma_1^F}{\Gamma_0^F}\,L_\perp + \frac{d_2^q}{\Gamma_0^F} \right) 
    \right] + {\cal O}(\alpha_s^2) \,,
\end{aligned}
\end{equation}
which is completely determined in terms of $F_{q\bar q}$. The two-loop coefficients $d_2^q$ and $\Gamma_1^F$ enter here already at next-to-leading order in $\alpha_s$. For a consistent resummation at NNLL order (or next-to-leading order in RG-improved perturbation theory), we need the one-loop expressions for the matching coefficients $C_V$ and $I_{i\leftarrow j}$, the two-loop expression for the exponent $F_{q\bar q}$, the two-loop expression for the anomalous dimension $\gamma^q$, and the three-loop cusp anomalous dimension and $\beta$ function. All of these ingredients are known.

Note also that, owing to the simple $q_T^2$ dependence of the resummed result (\ref{master}), it is trivial to perform the integral over transverse momentum required to calculate the cross section (\ref{fact3}) defined with a cut on transverse momentum.

\section{Asymptotic divergence and reorganized expansion}
\label{sec:asymp}

Despite the fact that it correctly resums all large logarithmic terms in the perturbative series, the elegant formula (\ref{master}) just derived is of limited practical use. The reason is a strong factorial divergence of the perturbative expansion coefficients resulting from terms in the functions $I_{i\rightarrow j}$ and $E_{q\bar q}$ of order $\big(\alpha_s L_\perp^2\big)^n$. To understand the origin of this effect, we recall from (\ref{beauty}) that {\em before\/} expansion in powers of $\alpha_s$ the hard-scattering kernels contain quadratic terms in $L_\perp$ in the exponent. The same is true for the ${\cal O}\big[\big(\alpha_s L_\perp^2\big)^n\big]$ terms in the kernels $I_{i\rightarrow j}$, which exponentiate as a consequence of the cusp logarithm in the evolution equation (\ref{Ievol}). Let us then consider, instead of (\ref{Fourier}), the Fourier integral
\begin{equation}\label{newFourier}
   \frac{1}{4\pi}\!\int\!d^2x_\perp\,e^{-iq_\perp\cdot x_\perp}\,
    e^{-\eta L_\perp - \frac14 a L_\perp^2} 
   = \frac{e^{-2\gamma_E}}{\mu^2}\!\int_{-\infty}^\infty\!d\ell\,
    J_0\Big(e^{\ell/2}\,b_0\,\frac{q_T}{\mu} \Big)\,
    e^{(1-\eta)\ell - \frac14 a\ell^2} 
   \equiv \frac{e^{-2\gamma_E}}{\mu^2}\,K\Big(\eta,a,\frac{q_T^2}{\mu^2}\Big) \,,
\end{equation}
where $b_0=2e^{-\gamma_E}$, and in the case at hand
\begin{equation}\label{adef}
   a = \frac{\alpha_s(\mu)}{2\pi} \left[ \Gamma_0^F + \eta_F(M^2,\mu)\,\beta_0 \right] .
\end{equation}
Some useful properties of the function $K(\eta,a,r)$ are summarized in Appendix~\ref{app:c}. The above definition is such that for $a=0$ we recover, up to a trivial factor, the result (\ref{Fourier}) with $n=0$:
\begin{equation}
   K(\eta,0,r) = r^{\eta-1} \frac{\Gamma(1-\eta)}{e^{2(\eta-1)\gamma_E}\,\Gamma(\eta)} \,.
\end{equation}

Keeping the quadratic term in the exponent vastly improves the convergence behavior of the Fourier integral. For $a=0$ (i.e., without the quadratic term) the integral on the left-hand side of (\ref{newFourier}) converges in the ultraviolet (for $x_T\to 0$) only if $\eta<1$, and for $\eta<\frac14$ its value must be defined by analytic continuation. For $a>0$, on the other hand, the integral converges for all values of $\eta$. It is then perhaps not surprising that any attempt to expand the Gaussian weight factor in a perturbative series leads to a badly behaved expansion. Indeed, writing the formal series
\begin{equation}
   K(\eta,a,r)\big|_{\rm exp} 
   = \sum_{n=0}^\infty\,\frac{1}{n!} \left( - \frac{a}{4} \right)^n
    \partial_\eta^{2n} K(\eta,0,r) 
   = \sum_{n=0}^\infty\,\frac{1}{n!} \left( - \frac{a}{4} \right)^n
    \partial_\eta^{2n}\,r^{\eta-1} 
    \frac{\Gamma(1-\eta)}{e^{2(\eta-1)\gamma_E}\,\Gamma(\eta)} \,,
\end{equation}
it is not difficult to see that the series is factorially divergent. To illustrate this point, we consider the special case where $r=1$ (corresponding to the default scale choice $\mu=q_T$) and $\eta$ is close to the critical value 1. One then has
\begin{equation}
   \frac{\Gamma(1-\eta)}{e^{2(\eta-1)\gamma_E}\,\Gamma(\eta)}
   = \frac{1}{1-\eta} - \frac{2\zeta_3}{3}\,(1-\eta)^2 - \frac{2\zeta_5}{5}\,(1-\eta)^4
    + \dots \,,
\end{equation}
and taking $2n$ derivatives of the leading term generates $(2n)!/(1-\eta)^{2n+1}$. A more careful analysis reveals that 
\begin{equation}\label{I1exp}
   K(\eta,a,1)\big|_{\rm exp} 
   = \sum_{n=0}^\infty\,\frac{(2n)!}{n!} \left( - \frac{a}{4} \right)^n
    \left[ \frac{1}{(1-\eta)^{2n+1}} - e^{-2\gamma_E} \right] 
    + \sum_{n=0}^\infty\,k_n\,a^n + {\cal O}(1-\eta) \,,
\end{equation}
where the coefficients $k_n$ do not exhibit the strong factorial growth of the terms in the first sum. While this series is badly divergent, the fact that it has alternating sign implies that it can be Borel-summed. We obtain
\begin{equation}\label{I1Borel}
\begin{aligned}
   K(\eta,a,1)\big|_{\rm Borel} 
   &= \sqrt{\frac{\pi}{a}} \left\{ e^{\frac{(1-\eta)^2}{a}} 
    \left[ 1 - \mbox{Erf}\left( \frac{1-\eta}{\sqrt{a}} \right) \right]
    - e^{-2\gamma_E+\frac1a} \left[ 1 - \mbox{Erf}\left( \frac{1}{\sqrt{a}} \right) 
    \right] \right\} \\
   &\quad\mbox{}+ \sum_{n=0}^\infty\,k_n\,a^n + {\cal O}(1-\eta) \,,
\end{aligned}
\end{equation}
where $\mbox{Erf}(x)$ is the error function. Note that the singularity at $\eta=1$ has disappeared after Borel summation. Expressions for the first few $k_n$ coefficients can be readily derived in terms of $\gamma_E$ and $\zeta_n$ values. Numerically, we find $k_0\approx 0.3152$, $k_1\approx 0.2431$, $k_2\approx-0.0747$, $k_3\approx-0.0509$, $k_4\approx 0.0446$, $k_5\approx-0.0043$, \dots. We have checked numerically that the Borel-summed result (\ref{I1Borel}) reproduces the original integral as defined in (\ref{newFourier}) precisely. 

The presence of factorially divergent terms in the perturbative expansion of the CSS formula for the resummed Drell-Yan cross section at small transverse momentum was first pointed out in \cite{Frixione:1998dw}. These authors considered this a problematic feature of $q_T$ resummation, which arises because the integral over $x_T$ in (\ref{Cdef}) includes the region $x_T\lesssim M^{-1}$, where the resummation formula is not justified. For the same reason, the authors of \cite{Bozzi:2003jy} proposed a modification of the formula in which one replaces $\ln\frac{x_T^2\mu^2}{b_0^2}\to\ln[(\frac{x_T^2}{b_0^2}+\frac{1}{M^2})\mu^2]$, thereby effectively introducing a UV cutoff on the $x_T$ integral. Note, however, that the quadratic term in the exponent of the Fourier integral (\ref{newFourier}) provides a Gaussian cutoff ensuring that $|L_\perp|\lesssim 1/\sqrt{a}$, which automatically excludes values $x_T\lesssim M^{-1}$.\footnote{For $\eta<1$ the integral is UV convergent even without the quadratic term, and contributions from the region $x_T<q_T^{-1}$ are power suppressed.} 
Our interpretation of the origin of the divergent terms is a different one. The asymptotic behavior of the perturbative series in gauge theories is a hint that perturbation theory by itself can only provide an incomplete description of nature. For instance, the appearance of infrared renormalon singularities -- factorially divergent terms related to running-coupling effects, whose Borel sum is ambiguous by power-suppressed amounts -- hints at the existence of non-perturbative power corrections, which must be added to the perturbative series in order to obtain a precise description of strong-interaction physics \cite{Beneke:1998ui}. In the present case, the appearance of a factorially divergent series indicates that some essential features of the transverse-momentum distribution in the region of small to moderate $q_T$ cannot be reproduced at any fixed order of perturbation theory, because they carry some genuinely non-perturbative information (even though this does not necessarily indicate a sensitivity to long-distance physics). It is well-known that for sufficiently large $M$ the intercept of the distribution $d\sigma/dq_T^2$ is calculable and depends in an intrinsically non-perturbative way on the coupling constant, $d\sigma/dq_T^2|_{q_T=0}\propto e^{{\rm const}/\alpha_s}/\sqrt{\alpha_s}$ \cite{Parisi:1979se,Collins:1984kg}. Our Borel-summed expression (\ref{I1Borel}) shares many of the features of the formula for the intercept. Only after Borel summation of the divergent series, which is possible due to its alternating sign behavior, we obtain a reliable description of the $q_T$ spectrum, which encodes its non-perturbative behavior in an appropriate way \cite{inprep}.

Let us now discuss the implications of these findings for the resummed Drell-Yan cross section. The main lesson is that in order to avoid an asymptotic growth of the perturbative expansion coefficients one should keep the quadratic terms in $L_\perp$ in the exponent of the Fourier integral (\ref{newFourier}). This gives rise to a new formula for the hard-scattering kernels in (\ref{Cdef}). Instead of the expression (\ref{master}), we obtain
\begin{equation}\label{newmaster}
\begin{aligned}
   C_{q\bar q\to ij}(z_1,z_2,q_T^2,M^2,\mu)
   &= \left| C_V(-M^2,\mu) \right|^2 \widehat I_{q\leftarrow i}(z_1,-\partial_\eta,\alpha_s)\,
    \widehat I_{\bar q\leftarrow j}(z_2,-\partial_\eta,\alpha_s) \\
   &\quad\times \widehat E_{q\bar q}(-\partial_\eta,\alpha_s,\eta_F)\,
    \frac{e^{-2\gamma_E}}{\mu^2}\,K\Big(\eta,a,\frac{q_T^2}{\mu^2} \Big) 
    \bigg|_{\eta=\eta_F(M^2,\mu)} \,, 
\end{aligned}
\end{equation}
where $a$ is given by (\ref{adef}), and the functions $\widehat I_{i\rightarrow j}$ and $\widehat E_{q\bar q}$ are obtained simply by dropping the ${\cal O}(\alpha_s L_\perp^2)$ terms in the expressions for $I_{i\rightarrow j}$ and $E_{q\bar q}$ in (\ref{Iresults}) and (\ref{beauty}), and likewise in higher orders. The integral $K(\eta,a,r)$ and its derivative with respect to $\eta$ can easily be evaluated numerically. We have not succeeded to derive a suitable analytic expression for this integral in the general case where $r\ne 1$ and $\eta$ is not close to 1.

\section{Comparison with the literature}
\label{sec:comp}

The standard formalism for transverse-momentum resummation has been developed in a seminal paper by Collins, Soper, and Sterman (CSS) \cite{Collins:1984kg}. According to this work, the resummed differential cross section at leading power can be written in the form
\begin{eqnarray}\label{CSS}
   \frac{d^3\sigma}{dM^2\,dq_T^2\,dy} 
   &=& \frac{4\pi\alpha^2}{3N_c M^2 s}\,
    \frac{1}{4\pi} \int\!d^2x_\perp\,e^{-iq_\perp\cdot x_\perp}
    \sum_q\,e_q^2\,\sum_{i=q,g} \sum_{j=\bar q,g}
    \int_{\xi_1}^1\!\frac{dz_1}{z_1} \int_{\xi_2}^1\!\frac{dz_2}{z_2} \nonumber\\
   &\times& \exp\left\{ - \int_{\mu_b^2}^{M^2}\!\frac{d\bar\mu^2}{\bar\mu^2}
    \left[ \ln\frac{M^2}{\bar\mu^2}\,A\big(\alpha_s(\bar\mu)\big)
    + B\big(\alpha_s(\bar\mu)\big) \right] \right\} \\
   &\times& \bigg[ C_{qi}\big(z_1,\alpha_s(\mu_b)\big)\,
    C_{\bar qj}\big(z_2,\alpha_s(\mu_b)\big)\,\phi_{i/N_1}(\xi_1/z_1,\mu_b)\,
    \phi_{j/N_2}(\xi_2/z_2,\mu_b) + (q,i\leftrightarrow\bar q,j) \bigg] , \nonumber
\end{eqnarray}
where $\mu_b=b_0/x_T$ is assumed to be in the perturbative domain. It is a straightforward exercise to work out the relations between the various objects in this formula and ours. We find
\begin{equation}\label{rels}
\begin{aligned}
   A\big(\alpha_s\big) 
   &= \Gamma_{\rm cusp}^F(\alpha_s) - \frac{\beta(\alpha_s)}{2}\,
    \frac{dg_1(\alpha_s)}{d\alpha_s} \,, \\
   B\big(\alpha_s\big) 
   &= 2\gamma^q(\alpha_s) + g_1(\alpha_s) - \frac{\beta(\alpha_s)}{2}\,
    \frac{dg_2(\alpha_s)}{d\alpha_s} \,, \\[2mm]
   C_{ij}\big(z,\alpha_s(\mu_b)\!\big)
   &= \left| C_V(-\mu_b^2,\mu_b) \right| I_{i\leftarrow j}\big(z,0,\alpha_s(\mu_b)\big) \,,
\end{aligned}
\end{equation}
where
\begin{equation}
\begin{aligned}
   g_1(\alpha_s) 
   &= F(0,\alpha_s) 
    = \sum_{n=1}^\infty\,d_n^q \left( \frac{\alpha_s}{4\pi} \right)^n , \\[-2mm]
   g_2(\alpha_s) 
   &= \ln\left| C_V(-\mu^2,\mu) \right|^2 
    = \sum_{n=1}^\infty\,e_n^q \left( \frac{\alpha_s}{4\pi} \right)^n .
\end{aligned}
\end{equation}
The one-loop coefficients are $d_1^q=0$ and
\begin{equation}
   e_1^q = C_F \left( \frac{7\pi^2}{3} - 16 \right) .
\end{equation}
The two-loop coefficient $d_2^q$ has been given in (\ref{d2resu}), while $e_2^q$ can be extracted from the results compiled in \cite{Becher:2006mr}, however it contributes to $B(\alpha_s)$ at ${\cal O}(\alpha_s^3)$ only. We have checked that the relations in (\ref{rels}) are compatible with our perturbative results.

Note that according to (\ref{rels}) the coefficient $A$ in the CSS formula differs from the cusp anomalous dimension starting at three-loop order, and the coefficient $B$ differs from the quark anomalous dimension $2\gamma^q$ starting at two-loop order.\footnote{The first relation in (\ref{rels}) can be found, in almost precisely this form, in equation (3.13) of \cite{Collins:1984kg}, from which it follows that $F_{q\bar q}(x_T^2,\mu)=-K(x_T\mu,\alpha_s)$ in the notation of that paper.}  
The first non-zero deviations are (here $A^{(n)}$ and $B^{(n)}$ denote the $n$-th order coefficients in the expansion in powers of $\alpha_s/(4\pi)$)
\begin{equation}\label{eq:A3B2}
   A^{(3)} = \Gamma_2^F + 2\beta_0 d_2^q \,, \qquad
   B^{(2)} = 2\gamma_1^q + d_2^q + \beta_0 e_1^q \,.
\end{equation}
The two-loop expression for $B(\alpha_s)$ was obtained a long time ago in \cite{Davies:1984sp}, while for gluon-initiated processes such as Higgs-boson production the corresponding coefficient was calculated in \cite{deFlorian:2001zd}. Using these results, we have derived the anticipated relation (\ref{d2resu}). Inserting the coefficients $d_2^{q,g}$ into (\ref{eq:A3B2}), we obtain the coefficient $A^{(3)}$, which up to now was the last missing ingredient for a full NNLL resummation of the $q_T$ spectrum. In the literature it is often assumed that $A^{(3)}=\Gamma_2^F$ (see e.\ g.\ \cite{Bozzi:2003jy,Balazs:2007hr}), which is true for soft gluon resummation, but our results show that for transverse-momentum resummation an extra contribution arises because of the collinear anomaly. Numerically, for the quark case with $n_f=5$, we find $\Gamma_2^F=239.2$ while $A^{(3)}=-413.7$, so the extra term is much larger than the contribution from the cusp anomalous dimension and has opposite sign. It will be interesting to see how this changes the numerical predictions for the spectrum. Note also that, due to Casimir scaling, in the gluon case a similar situation but with larger coefficients occurs, and we find $\Gamma_2^A=538.2$ while $A^{(3)}=-930.8$. 

CSS also derive a formula for the resummed cross section at small $q_T$ in terms of transverse-position dependent PDFs $\overline{\cal P}_{i/N}(\xi,x_T)$, which is equivalent to our result (\ref{Bfact}) once we identify
\begin{equation}\label{CSSmatch}
   \overline{\cal P}_{i/N}(\xi,x_T)
   = \left| C_V(-\mu_b^2,\mu_b) \right| B_{i/N}(\xi,x_T^2,\mu_b) \,.
\end{equation}
The definition of the function $\overline{\cal P}_{i/N}(\xi,x_T)$ proceeds in two steps. One first solves a differential equation governing the $\zeta$ dependence of the gauge-dependent distribution function $\tilde{\cal P}_{i/N}(\xi,x_T,\mu;\zeta)$ mentioned earlier in our discussion in Section~\ref{sec:collinearanomaly}. This introduces a gauge-independent boundary function $\hat{\cal P}(\xi,x_T,\mu)$ \cite{Collins:1981uk}. The function $\overline{\cal P}_{i/N}(\xi,x_T)$ is obtained by multiplying this function with an $x_T$-dependent factor that cancels its scale dependence \cite{Collins:1981va}. Like in our case, no operator definition for the final transverse PDF $\overline{\cal P}_{i/N}$ is provided.

While we obtained the closed-form expressions (\ref{CVsol}), (\ref{master}), and (\ref{newmaster}) for the cross section directly in momentum space, the CSS formula (\ref{CSS}) involves a Fourier integral over $x_T$. The inherent scale choice $\mu=\mu_b=b_0/x_T$ eliminates all $L_\perp$ logarithms and thus automatically resums the factorially divergent terms discussed in Section \ref{sec:asymp}. However, since the integrand involves $\alpha_s(\mu_b)$, the integration hits the Landau pole in the running coupling, so that a prescription is needed to regularize the integral. In practical applications, the integration is cut off at large $x_T$ values. To account for the missing long-distance contributions a non-perturbative model function is used. For $q_T$ in the perturbative domain these contributions are formally power suppressed, but it is irritating that an explicit prescription for how to deal with them is needed even for $q_T$ values deep in the perturbative regime. Special care has to be taken in order not to induce unphysical power corrections in the cut-off procedure \cite{Beneke:1995pq}. The explicit cut-off also makes it difficult to perform the matching to fixed-order computations, since resummation effects persist even when the CSS formula is evaluated at large $q_T\sim M$, as discussed in \cite{Ellis:1997ii}. All of these complications are absent in our resummed results (\ref{CVsol}), (\ref{master}), and (\ref{newmaster}). Our expressions are given directly in momentum space and do not involve a Landau-pole singularity. In the spirit of effective field theory, we never perform scale setting inside integrals over the running coupling $\alpha_s(\mu)$. Instead, the scales are chosen such that the integrated result is free of large logarithms. The matching onto fixed-order computations is completely trivial, since our analytic result (\ref{master}) can easily be reexpanded in powers of a fixed coupling $\alpha_s(\mu)$. Moreover, in contrast to the CSS formula, the resummation switches itself off adiabatically when $\mu\sim q_T$ approaches $M$, since all the logarithms become small and the RG evolution stops when these scales become equal. 

The advantages of performing the resummation in $q_T$ rather than in $x_T$ space were emphasized a long time ago in \cite{Ellis:1997ii}, where an improved leading-log formula extending the work of \cite{DDT} was derived, which resums all logarithmically-enhanced terms to two-loop order. We have checked that this formula is consistent with our exact result (\ref{master}) provided the two-loop modification of the $\tilde B^{(2)}$ shown in equation (34) of \cite{Ellis:1997ii} is included. Note that the approximation adopted in that paper misses terms of order $(\alpha_s L)^n L^{n-4}$ for $n\ge 3$ (denoting $L=\ln(M^2/q_T^2)$), which for $\alpha_s L=O(1)$ become progressively enhanced for increasing $n$. In our approach, all such enhanced terms are resummed.

Transverse-momentum resummation has been studied earlier using SCET in \cite{Gao:2005iu,Idilbi:2005er} and more recently in \cite{Mantry:2009qz}. These papers solve the RG evolution equation of the hard function $|C_V|^2$ in a way that is equivalent to (\ref{CVsol}). The exponentiation of the additional large logarithms of $q_T^2/M^2$, which arise from the transverse PDFs due to the collinear anomaly, is however not addressed in any of these works. Because of this, the relations derived in \cite{Gao:2005iu,Idilbi:2005er} between the effective-theory results and the traditional CSS approach miss the extra contributions proportional to $g_1$ in (\ref{rels}) and hence are only valid to next-to-leading logarithmic order. In \cite{Mantry:2009qz} the resummation of large logarithms was studied in momentum space, similar to our approach. However, since their result is given in terms of a multi-dimensional convolution integral it is more complicated to implement in practice. More importantly, the fact that the large logarithms arising from the collinear anomaly (which would have to appear in the matching conditions for the jet and soft functions in that paper or in the convolutions of these objects) were neither addressed nor resummed implies that the resummation formulae derived in \cite{Mantry:2009qz} will have to be modified already at NLL order (with the usual counting of logarithms in the exponent). The reason is that they do not accomplish the resummation of the $\eta_F$-dependent terms in (\ref{master}).

At first sight, the papers \cite{Gao:2005iu,Idilbi:2005er,Mantry:2009qz} avoid dealing with the collinear anomaly by keeping power-suppressed terms in the computation of the soft function and of the matching of the transverse onto the usual PDFs. In particular, the multipole expansion of the soft function is not performed, and for this reason the soft function (\ref{Wsoft}) survives in the final factorization formula. While keeping higher-order terms regulates the light-cone singularities, care must to be taken to avoid double counting (this requires so-called zero-bin subtractions). Also, if the multipole expansion is not performed consistently, the soft and beam functions depend on several scales, such that their computation is not only much more cumbersome than in our case, but also gives rise to extra logarithms, which would need to be resummed. No attempt is made in \cite{Gao:2005iu,Idilbi:2005er,Mantry:2009qz} to derive the corresponding evolution equations.

\section{Conclusions} 
\label{sec:con}

We have analyzed the production of high-mass Drell-Yan pairs at low transverse momentum $q_T$ using effective field-theory methods. Since the classical paper of Collins, Soper, and Sterman from 1985, it is known how to resum the large perturbative logarithms of $q_T$ over the invariant mass of the lepton pair which arise in this kinematical region. Despite this, certain aspects of the factorization properties of the cross section at small $q_T$ have continued to be puzzling. The difficulties arise because the cross section factors into a product of transverse-position dependent PDFs, whose naive definition leads to inconsistencies. Many different ``improved'' definitions for these objects were proposed over the past thirty years, but none of them seems entirely satisfactory.

In this paper we have shown that while individual $x_T$-dependent PDFs are not well defined without additional regulators, their product is. However, this product depends on the large momentum transfer $q^2\gg 1/x_T^2$ of the underlying hard-scattering process. This anomaly arises because the regulators needed to define the individual PDFs in SCET necessarily break the naive factorization of the effective Lagrangian into two independent collinear sectors. Similar to other anomalies, the breaking of factorization has very specific properties, which imply that the dependence on the momentum transfer exponentiates. Only after the $q^2$ dependence associated with the anomaly is accounted for, the remainder factors into a product of two functions, each dependent on the transverse separation and the fraction of the large momentum carried by the leading parton. These functions provide an operational definition (though not an operator definition) of transverse-position dependent PDFs, whose evolution equation is solved exactly and is known explicitly at three-loop order. However, it is an open question whether the anomalous $q^2$ dependence is truly universal and process independent. We note that at large $x_T\sim 1/\Lambda_{\rm QCD}$ the exponent of the $q^2$ dependence from the collinear anomaly becomes a non-perturbative quantity, so that the dependence on the large momentum transfer $q^2\gg\Lambda_{\rm QCD}^2$ is no longer perturbatively calculable. For small transverse separation $x_T\ll 1/\Lambda_{\rm QCD}$, on the other hand, the transverse PDFs can be matched onto standard PDFs, and their dependences on $x_T$ and $q^2$ are calculable in perturbation theory. We have performed the necessary matching calculation at one-loop order and derived the $q^2$ dependence due to the anomaly at the two-loop level.

As a result of our factorization analysis, we are able to obtain the first, closed analytic expression for the cross section at small $q_T\ll M$ directly in momentum space, which is free from large perturbative logarithms. In addition to logarithms arising from the evolution of the hard function to lower scales, which we resum by RG evolution and which also arise in soft-gluon resummation, our result contains a second source of large logarithms due to the collinear anomaly. A crucial result of our analysis is that these additional logarithms exponentiate. We have also shown that, besides parametrically large logarithms, the perturbative series for the transverse-momentum distribution contains terms exhibiting a strong factorial growth at higher orders in $\alpha_s$, which must be summed in order to obtain reliable results. We have explained the origin of these terms and shown how they can be accounted for to all orders in perturbation theory.

Our momentum-space expression for the resummed cross section offers a number of advantages over the traditional formalism. Since we have an analytic result for the resummed cross section, we can easily reexpand it to match to fixed-order results, while the matching is nontrivial in the traditional approach. Our resummation automatically switches off as $q_T^2$ approaches $q^2$. In addition, our result is free from unphysical Landau-pole singularities, so that no explicit prescription for the non-perturbative regime is needed as long as $q_T\gg \Lambda_{\rm QCD}$. We have derived all the necessary input to perform resummation at next-to-next-to-leading logarithmic accuracy and as a by-product have obtained the three-loop coefficient $A^{(3)}$, which is needed in the traditional approach to achieve this accuracy. A phenomenological analysis of electroweak boson production using our formalism will be presented elsewhere.

\vspace{0.2cm}
{\em Acknowledgments:\/}
We are grateful to Stefano Catani, Daniel de Florian, Pavel Nadolsky, and Dave Soper for useful discussions. We would like to thank the Aspen Center for Physics, where part of this research was performed. The research of M.N.\ is supported in part by a Jensen Professorship from the Klaus Tschira Foundation, BMBF grant 05H09UME, DFG grant NE 398/3-1, and the Research Centre {\em Elementary Forces and Mathematical Foundations\/}. T.B.\ is supported in part by SNSF and ``Innovations- und Kooperationsprojekt C-13'' of SUK.

\newpage
\begin{appendix}

\section{Generalization to $\bm{W}$ and $\bm{Z}$ production}
\label{app:b}
\renewcommand{\theequation}{A\arabic{equation}}
\setcounter{equation}{0}

The generalization of our results from the photon-induced Drell-Yan process discussed in the paper to $W$ and $Z$ production is straightforward. To obtain the double differential cross section $d^2\sigma/dq_T^2\,dy$ from (\ref{Bfact}) and (\ref{fact1}), one needs to change the prefactor according to
\begin{equation}
   \frac{4\pi\alpha^2}{3N_c M^2 s} \to \frac{4\pi^2\alpha}{N_c\,s}
\end{equation}
and insert the proper charge factors. For the $Z$ boson one must replace
\begin{equation}
   \sum_q\,e_q^2 \to \sum_q\,\frac{|g_L^q|^2+|g_R^q|^2}{2}
   = \sum_q\,\frac{\big( 1 - 2|e_q|\sin^2\theta_W \big)^2 + 4 e_q^2\sin^4\theta_W}%
                  {8\sin^2\theta_W\cos^2\theta_W} \,,
\end{equation}  
where $\theta_W$ is the weak mixing angle. Since the $W$ bosons have flavor-changing couplings, to sum over flavors must be replaced by a double sum over individual quark and anti-quark flavors, $q$ and $q'$. Only left-handed currents appear in this case. The relevant coupling for a $W^-$ boson produced in the annihilation of an anti-up and a down quark is
\begin{equation}
   \sum_q\,e_q^2 \to \sum_{q,q'}\,\frac{|g_L^{q'q}|^2}{2}
   = \sum_{q,q'}\,\frac{|V_{q'q}|^2}{4\sin^2\theta_W} \,,
\end{equation}
where $V_{q'q}$ are elements of the quark mixing matrix.

\section{RG evolution}
\label{app:a}
\renewcommand{\theequation}{B\arabic{equation}}
\setcounter{equation}{0}

Here we give the perturbative expansions of the functions $S$ and $a_\Gamma$ defined in (\ref{RGEsols}), working consistently at next-to-leading order in RG-improved perturbation theory. At this order we need to keep terms through $O(\alpha_s)$ in the final expressions. The result for $a_\Gamma$ is given by \cite{Neubert:2004dd}
\begin{equation}\label{asol}
   a_\Gamma(\nu,\mu)
   = \frac{\Gamma_0^F}{2\beta_0} \left[ \,\ln\frac{\alpha_s(\mu)}{\alpha_s(\nu)}
    + \left( \frac{\Gamma_1^F}{\Gamma_0^F} - \frac{\beta_1}{\beta_0} 
    \right) \frac{\alpha_s(\mu) - \alpha_s(\nu)}{4\pi} + \dots \right] .
\end{equation}
A similar expressions, with the coefficients $\Gamma_i^F$ replaced by $\gamma_i^V$, holds for the function $a_{\gamma^q}$. The expression for the Sudakov exponent $S$ is more complicated. It reads \cite{Neubert:2004dd}
\begin{equation}
\begin{aligned}
   S(\nu,\mu) 
   &= \frac{\Gamma_0^F}{4\beta_0^2}\,\Bigg\{
    \frac{4\pi}{\alpha_s(\nu)} \left( 1 - \frac{1}{r} - \ln r \right)
    + \left( \frac{\Gamma_1^F}{\Gamma_0^F} - \frac{\beta_1}{\beta_0}
    \right) (1-r+\ln r) + \frac{\beta_1}{2\beta_0} \ln^2 r \\
   &\hspace{1.6cm}\mbox{}+ \frac{\alpha_s(\nu)}{4\pi} \Bigg[ 
    \left( \frac{\beta_1\Gamma_1^F}{\beta_0\Gamma_0^F} 
    - \frac{\beta_2}{\beta_0} \right) (1-r+r\ln r)
    + \left( \frac{\beta_1^2}{\beta_0^2} 
    - \frac{\beta_2}{\beta_0} \right) (1-r)\ln r \\
   &\hspace{3.4cm}
    \mbox{}- \left( \frac{\beta_1^2}{\beta_0^2} 
    - \frac{\beta_2}{\beta_0}
    - \frac{\beta_1\Gamma_1^F}{\beta_0\Gamma_0^F} 
    + \frac{\Gamma_2^F}{\Gamma_0^F} \right) \frac{(1-r)^2}{2} \Bigg] + \dots \Bigg\} \,,
\end{aligned}
\end{equation}
where $r=\alpha_s(\mu)/\alpha_s(\nu)$. Whereas the two-loop anomalous dimensions and $\beta$-function are required in (\ref{asol}), the expression for $S$ also involves the three-loop coefficients $\Gamma_2^F$ and $\beta_2$.

The relevant expansion coefficients of the cusp anomalous dimension $\Gamma_{\rm cusp}^F$ to three-loop order are $\Gamma_0^F=4C_F$ and \cite{Moch:2004pa}
\begin{equation}
\begin{aligned}
   \frac{\Gamma_1^F}{\Gamma_0^F} 
   &= \left( \frac{67}{9} - \frac{\pi^2}{3} \right) C_A - \frac{20}{9}\,T_F n_f \,, \\
   \frac{\Gamma_2^F}{\Gamma_0^F} 
   &= C_A^2 \left( \frac{245}{6} - \frac{134\pi^2}{27} + \frac{11\pi^4}{45}
    + \frac{22}{3}\,\zeta_3 \right) 
    + C_A T_F n_f  \left( - \frac{418}{27} + \frac{40\pi^2}{27}
    - \frac{56}{3}\,\zeta_3 \right) \\
   &\quad\mbox{}+ C_F T_F n_f \left( - \frac{55}{3} + 16\zeta_3 \right) 
    - \frac{16}{27}\,T_F^2 n_f^2 \,.
\end{aligned}
\end{equation}
Due to Casimir scaling, these ratios are the same in any representation of the gauge group. The coefficients of the quark anomalous dimension $\gamma^q$ to two-loop order read \cite{Becher:2006mr,Becher:2009qa}
\begin{eqnarray}
   \gamma_0^q &=& -3 C_F \,, \\
    \gamma_1^q  &=& C_F^2 \left( -\frac{3}{2} + 2\pi^2 - 24\zeta_3 \right)
    + C_F C_A \left( - \frac{961}{54} - \frac{11\pi^2}{6} 
    + 26\zeta_3 \right)
    + C_F T_F n_f \left( \frac{130}{27} + \frac{2\pi^2}{3} \right) . \nonumber
\end{eqnarray}
The three-loop coefficient $\gamma_2^q$ can be found in the same papers. Finally, the expansion coefficients for the QCD $\beta$-function to three-loop order are
\begin{equation}
\begin{aligned}
   \beta_0 &= \frac{11}{3}\,C_A - \frac43\,T_F n_f \,, \\
   \beta_1 &= \frac{34}{3}\,C_A^2 - \frac{20}{3}\,C_A T_F n_f
    - 4 C_F T_F n_f \,, \\
   \beta_2 &= \frac{2857}{54}\,C_A^3 + \left( 2 C_F^2
    - \frac{205}{9}\,C_F C_A - \frac{1415}{27}\,C_A^2 \right) T_F n_f
    + \left( \frac{44}{9}\,C_F + \frac{158}{27}\,C_A 
    \right) T_F^2 n_f^2 \,.
\end{aligned}
\end{equation}

We finally present the exact solutions to the RG equations for the exponent $F_{q\bar q}$ and the function $B_{q/N}$ given in (\ref{Bevol}). The solution to the first equation reads 
\begin{equation}
   F_{q\bar q}(x_T^2,\mu) = - 2a_\Gamma(\mu_0,\mu) + F_{q\bar q}(x_T^2,\mu_0)
   = - 2a_\Gamma(\mu_b,\mu)
    + \sum_{n=1}^\infty\,d_n^q \left( \frac{\alpha_s(\mu_b)}{4\pi} \right)^ n \,,
\end{equation}
where $\mu_0$ is an arbitrary initial scale of order $q_T$ or $x_T^{-1}$, and in the last step we have made the special choice $\mu_0=\mu_b=2e^{-\gamma_E}/x_T$. The solution to the second equation is 
\begin{equation}
   B_{i/N}(z,x_T^2,\mu)
   = \exp\left[ - 2S(\mu_0,\mu) + 2a_{\gamma^q}(\mu_0,\mu) \right] 
    \left( \frac{x_T^2\mu_0^2}{4e^{-2\gamma_E}} \right)^{-2a_\Gamma(\mu_0,\mu)}
    B_{i/N}(z,x_T^2,\mu_0) \,.
\end{equation}
As before, a particularly convenient choice is to set $\mu_0=\mu_b$. This leads to 
\begin{equation}
   B_{q/N}(z,x_T^2,\mu)\equiv e^{h_{q\bar q}(x_T^2,\mu)}\,\overline{B}_{q/N}(z,x_T^2) \,,
\end{equation} 
where $h_{q\bar q}(x_T^2,\mu)=-2S(\mu_b,\mu)+2a_{\gamma^q}(\mu_b,\mu)$, while $\overline{B}_{q/N}(z,x_T^2)\equiv B_{q/N}(z,x_T^2,\mu_b)$ is RG invariant. This is the function that, up to a factor, is equal to the distribution function $\overline{\cal P}_{q/N}(z,x_T)$ introduced in \cite{Collins:1981va}, see (\ref{CSSmatch}). For $\mu\sim\mu_b$ one can construct a fixed-order expression for the exponent $h_{q\bar q}$ as a polynomial in $L_\perp$. At two-loop order we obtain
\begin{equation}
   h_{q\bar q}(x_T^2,\mu) = \sum_{n=1}^\infty\,h_n^q(L_\perp)
    \left( \frac{\alpha_s}{4\pi} \right)^n ,
\end{equation}
with   
\begin{equation}
   h_1^q(L_\perp) = \frac{\Gamma_0^F}{4}\,L_\perp^2 - \gamma_0^q\,L_\perp \,,
    \qquad
   h_2^q(L_\perp) = \frac{\beta_0\Gamma_0^F}{12}\,L_\perp^3 + 
    \left( \frac{\Gamma_1^F}{4} - \frac{\beta_0\gamma_0^q}{2} \right) L_\perp^2
    - \gamma_1^q\,L_\perp \,.
\end{equation}

\section{Properties of the function $\bm{K(\eta,a,r)}$}
\label{app:c}
\renewcommand{\theequation}{C\arabic{equation}}
\setcounter{equation}{0}

Explicit expressions for the Fourier integral can be obtained in the limits of large $a$ or small $r$. We obtain
\begin{equation}\label{largea}
   K(\eta,a,r) = \sqrt{\frac{4\pi}{a}} J_0(b_0\sqrt{r}) + {\cal O}(a^{-3/2}) \,;
    \qquad a\gg 1 \,,
\end{equation}
and
\begin{equation}
   K(\eta,a,r) = \sqrt{\frac{4\pi}{a}} \left[ e^{\frac{(1-\eta)^2}{a}}
    - e^{-2\gamma_E}\,r\,e^{\frac{(2-\eta)^2}{a}} + {\cal O}(r^2) \right] ;
     \qquad r\ll 1 \,.
\end{equation}

Noting that a rescaling of the ratio $q_T/\mu$ in the argument of the Bessel function in (\ref{newFourier}) can be compensated by a shift of the integration variable $\ell$, we find that
\begin{equation}
   K(\eta,a,r) = r^{\eta-1-\frac{a}{4}\ln r}\,\bar K\Big(\eta-\frac{a}{2}\ln r,a\Big) \,,
\end{equation}
where $\bar K(s,a)\equiv K(s,a,1)$. This function obeys the partial differential equation
\begin{equation}
   \left( \frac{\partial^2}{\partial s^2} + 4\,\frac{\partial}{\partial a} \right)
   \bar K(s,a) = 0 \,,
    \qquad
   \bar K(s,0) = \frac{\Gamma(1-s)}{e^{2(s-1)\gamma_E}\,\Gamma(s)} \,.
\end{equation}

The Borel image of the divergent part of the series (the first sum) in (\ref{I1exp}) is
\begin{equation}
   B_{\rm div}(u) 
   = \frac{1}{1-\eta} \left[ \delta(u) - \frac{1-\eta}{2\left(u+(1-\eta)^2\right)^{3/2}} 
    \right]
    - e^{-2\gamma_E} \left[ \delta(u) - \frac{1}{2\left(1+u\right)^{3/2}} \right] ,
\end{equation}
and performing the Borel integral $\int_0^\infty\!du\,B_{\rm div}(u)\,e^{-u/a}$ yields the expression shown in (\ref{I1Borel}). Note also that relation (\ref{largea}) implies that for large $a$ the non-divergent terms in the series must obey the sum rule
\begin{equation}
   \sum_{n=0}^\infty\,k_n\,a^n 
   = \sqrt{\frac{\pi}{a}}\,\Big[ 2 J_0(b_0) - 1 + e^{-2\Gamma_E} \Big] + {\cal O}(a^{-3/2})
   \approx \frac{1.2988}{\sqrt a} \,.
\end{equation}

\end{appendix}

\end{document}